\documentclass[useAMS,usenatbib]{mn2e}
\usepackage{graphicx}

\def\lesssim{\mathrel{\hbox{\rlap{\hbox{\lower4pt\hbox{$\sim$}}}\hbox{$<$}}}}
\def\gtrsim{\mathrel{\hbox{\rlap{\hbox{\lower4pt\hbox{$\sim$}}}\hbox{$>$}}}}

\def\gtrsim{\mathrel{\hbox{\rlap{\hbox{\lower4pt\hbox{$\sim$}}}\hbox{$>$}}}}

\usepackage{lscape}

\begin{document}

\title[NIRSPEC survey]{
The Emission Line Properties of Gravitationally-lensed 1.5$<z<$5 Galaxies}
\author[Richard et al.]{
 \parbox[h]{\textwidth}{
Johan Richard$^{1,2}$\thanks{E-mail:
johan.richard@durham.ac.uk}, Tucker Jones$^{3}$, 
Richard Ellis$^{3}$, Daniel P. Stark$^{4}$, Rachael Livermore$^{2}$, Mark Swinbank$^{2}$}
\vspace{6pt}\\
$^{1}$Dark Cosmology Centre, Niels Bohr Institute, University of
Copenhagen, Juliane Maries Vej 30, 2100 Copenhagen, Denmark\\
$^{2}$Institute for Computational Cosmology, Department of Physics, Durham University,
South Road, Durham, DH1 3LE\\
$^{3}$Astronomy Department, California Institute of Technology, MC105-24, Pasadena, CA 91125, USA \\
$^{4}$Institute of Astronomy, University of Cambridge, Madingley Road, Cambridge CB3 0HA\\
}

\date{Accepted 2010 December 6. Received 2010 November 29; in original form 2010 October 29}

\pagerange{\pageref{firstpage}--\pageref{lastpage}} \pubyear{2009}

\maketitle

\label{firstpage}

\begin{abstract}
We present and analyse near-infrared spectroscopy for a sample of 28 gravitationally-lensed
star-forming galaxies in the redshift range $1.5<z<5$, observed mostly with the Keck II telescope. With 
typical magnifications of $\simeq$1.5-4 magnitudes, our survey provides a valuable 
census of star formation rates, gas-phase metallicities and dynamical masses for a representative 
sample of low luminosity galaxies seen at a formative period in cosmic history.  We find less evolution 
in the mass-metallicity relation compared to earlier work that focused on more luminous systems
with $z\sim2-3$, especially in the low mass ($\sim10^9$ M$_\odot$) where our sample is $\sim0.25$ dex 
more metal-rich. We interpret this offset as a result of the lower star formation rates (typically a factor of $\sim$10 
lower) for a given stellar mass in our sub-luminous systems. Taking this effect into account, we conclude our objects 
are consistent with a fundamental metallicity relation recently proposed from unlensed observations. 
\end{abstract}


\section{Introduction}
The period corresponding to the redshift range $2<z<4$ is a formative one in the history of star-forming
galaxies. During this era, mass assembly proceeds at its fastest rate and the bulk of the metals
that establish present-day trends are likely manufactured. Through comprehensive multi-wavelength
surveys, much has been learned about the demographics of galaxies during this period 
\citep{Shapley03,VanDokkum,Chapman05} as summarized in recent global measures \citep{Hopkins,Ellis08}.

Attention is now focusing on the detailed properties of selected star-forming sources in this
redshift range. Integral field unit (IFU) spectrographs aided by adaptive optics correction on ground-based 
telescopes have delivered resolved velocity fields for various $z>2$ galaxy samples revealing
systemic rotation in a significant subset \citep{Genzel06,Genzel08,FS06,FS09,
Law07,Law09,Stark08,Jones}. Near-infrared spectroscopy, sampling rest-frame
optical nebular emission lines, defines the ongoing star formation rate and the gas-phase metallicity
\citep{Erb06,Mannucci}. Metallicity has emerged as a key parameter since it
measures the fraction of baryonic material already converted into stars. Quantitative measures
can thus be used to test feedback processes proposed to regulate star formation during an important 
period in cosmic history.

A fundamental relation underpinning such studies is the mass-metallicity relation first noted
locally by \citet{Lequeux} and recently quantified in the SDSS survey by \citet{Tremonti04}.
The oft-quoted explanation for the relation invokes star-formation driven outflows, e.g.
from energetic supernovae,  which have a larger effect in low mass galaxies with weaker
gravitational potentials. However, other effects may enter, particularly at high redshifts where
star formation timescales and feedback processes and their mass dependence likely differ.  

Motivated by the above, much observational effort has been invested to measure {\it evolution} in 
the mass-metallicity relation with redshift. The relationship has been defined using galaxy
samples extending beyond $z\simeq1$ \citep{Lamareille09,PM09} 
and $z\simeq2$ \citep{Erb06,Halliday, Hayashi}. Most recently,
\citet{Mannucci} have studied the properties of 10 Lyman-break galaxies (LBGs) 
at $z\simeq3$. These pioneering surveys have demonstrated clear evolution with
metallicities that decrease at earlier times for a fixed stellar mass. 

Inevitably as one probes to higher redshift, it becomes progressively harder to maintain
a useful dynamic range in the stellar mass and galaxy luminosity. In the case
of most distant studies (e.g. \citealt{Mannucci}), only with long integrations can
the mass-metallicity relation be extended down to masses of $10^9\,M_\odot$,
Samples defined via searches through gravitational lensing clusters are a much
more efficient probe of this important low mass regime. LBGs lensed by massive foreground 
clusters can be magnified by 2-3 magnitudes thereby probing intrinsically less massive 
systems. Initial results using this technique have been presented
for small samples by  \citet{Lemoine03}, \citet{Hainline} and \citet{Bian}.

As part of a long-term program to determine the resolved dynamical properties of
sub-luminous $z>2$ galaxies, we identified a large sample ($\simeq$30) of 
gravitationally-lensed systems with $z>$1.5 in the HST archive \citep{Sand05,Smith05,Richard10b} 
and embarked upon a systematic spectroscopic survey with the Keck II 
Near Infrared Spectrograph (NIRSPEC) to determine their emission line characteristics. 
The initial motivation was to use the high efficiency of NIRSPEC to screen each target 
prior to more detailed follow-up with IFU spectrographs sampling the star formation rate 
and velocity field across each source (\citealt{Jones}, Livermore et al. in preparation). 
However, a further product of this extensive spectroscopic survey is detailed information
on the star-formation rate, emission line ratios, and line widths for a large sample
of lensed $z>1.5$ galaxies. Our eventual sample comprises 28 objects including 5 
from the literature mentioned above. The goal of this paper is thus to utilize this
sample to extend studies of the mass-metallicity relation and related
issues to more representative lower luminosity galaxies at early times.

The paper is structured as follows. Section 2 introduces our sample and discusses
the various NIRSPEC observations and their reductions as well as associated
Spitzer data necessary to derive stellar masses. Section 3 discusses the
mass-metallicity relation and the relationship between dynamical mass and 
stellar mass noting that a subset of our sample has more detailed resolved
data (\citealt{Jones,Jones10b}). We discuss the implications of our results in the context
of measurements made of more luminous systems in Section 4. 

Throughout the paper, we assume a $\Lambda$-CDM cosmology with 
$\Omega_\Lambda$=0.7, $\Omega_m$=0.3 and $h=0.7$. For this cosmology and 
at the typical redshift $z\sim2.5$ of our sources, 1\arcsec\ on sky corresponds to 
$\sim$8.2 kpc. All magnitudes are given in the AB system.

\section{Observations and Data Reduction}

\subsection{Lensed Sample}

\begin{figure*}
\include{finders2}
\caption{\label{finders}Thumbnail HST images (V band) for each source targeted 
for near-infrared spectroscopy, showing the orientation of the corresponding 
NIRSPEC long slit. Targets are marked where there is otherwise ambiguity.}
\end{figure*}

We selected a sample of lensed galaxies for NIRSPEC follow-up 
using criteria similar to those used for for near-infrared spectroscopy
of LBGs (e.g. \citealt{Erb06}) but extended to lower intrinsic luminosities
after correction for the lensing magnification. The relevant criteria are:

\begin{itemize}

\item{Availability of optical data from the Hubble Space Telescope indicating a 
prominent rest-frame UV continuum  with $V<24$}

\item{Spectroscopic redshift $z>1.5$ derived from the literature or as part of our Keck spectroscopic campaign 
(\citealt{Richard07,Richard09,Richard10b}, Richard et al. 2010c in preparation)}

\item{An areal magnification factor $\mu \gtrsim1.5$ mag provided by the foreground lensing cluster for which 
a well-constrained mass model enables a good understanding of the associated errors (
e.g. \citealt{Richard10b}, Appendix A)}

\item{Emission lines predicted to lie in an uncontaminated region of the near-infrared night sky spectrum}
\end{itemize}

The application of these criteria generated a list of $\sim$ 50 arcs for further follow-up. We summarize in 
Table \ref{targets} the 23 sources drawn from this master list for which we were able to measure significant 
emission line fluxes. We have augmented this sample with ISAAC archival data and data from the literature for 
5 other targets  (see Sect. \ref{literature}). In total, we consider data for 28 lensed sources spanning the 
redshift range 1.5$<z<$4.86.

\subsection{Near-infrared spectroscopic data}
\subsubsection{NIRSPEC observations and data reduction}

The bulk of the spectroscopic survey was conducted with the NIRSPEC spectrograph
(\citealt{nirspec}) on the Keck II telescope in its low resolution mode during 
7 observing runs (Table \ref{runs}).  A 42 $\times$0.76\arcsec long slit was oriented along 
the major axis of each object, usually the direction of the highest magnification, in order to 
maximize the line fluxes (Fig. \ref{finders}). At this resolution, a different wavelength setup 
was selected for each target (filters N1 to N7, corresponding to the $z'$ to $K$  bands) for 
each group of lines ([O{\sc ii}], H$\beta$+[O{\sc iii}]$_{\lambda\lambda4959,5007}$, H$\alpha$+[N{\sc ii}]+[S{\sc ii}]). 

A major advantage of targeting lensed sources is the ability to survey a large sample
of low luminosity sources in an economic amount of observing time.
We typically undertook 2 to 4 dithered exposures of 300 - 600 secs each, depending on the 
magnitude of the source and the sky levels in a given band. We used a three point dithering pattern 
with offsets larger than the size of the object along the slit. Standard stars were used as flux calibrators. 
Simultaneously with the NIRSPEC integrations, we took  a series of 4-8 short exposures using the 
slit viewing camera, SCAM, in order to monitor the seeing and slit alignment with the object.

The data was reduced using IDL scripts following the procedure described in more detail  
in \citet{Stark07} and \citet{Richard08}. Although the 2D spectrum is distorted on the
detector, sky subtraction, wavelength and flux calibration were accomplished in the distorted 
frame, thereby mitigating any deleterious effect  of resampling (see \citealt{Kelson} for more details). 

In comparison with the general procedure applied by \citet{Stark07} for point sources, we took
special care to prevent sky over-subtraction for our bright and extended sources.

\subsubsection{Archival, literature and IFU data}
\label{literature}

A modest amount of additional data on lensed galaxies is available in the literature, in 
the archive, or through new IFU data. To date, the Keck survey described above represents the major advance.

We included data from the ISAAC instrument on the ESO VLT on 
2 lensed galaxies at $z$=1.9 \citep{Lemoine03}. Additional NIRSPEC data has
been published for two suitable objects by \citet{Hainline}. Finally, we retrieved 
ISAAC archival data for the giant arc in Cl2244 \citep{Hammer89} 
at $z=2.24$ which was studied by Lemoine-Busserolle et al. (2004). 
These archival J, H and K-band spectra have been reduced with standard IRAF scripts, 
following the procedure used in \citet{Richard03}. These 5 additional sources are listed 
separately in Table \ref{targets}.

Together with the IFU data presented in \citet{Jones}, three of the targets described above (the 
cosmic eye, A1835 and MACS0451) were observed with the SINFONI integral field spectrograph 
on the VLT between May 2009 and September 2010 as part of program 083.B-0108.  In all 
three cases we used a 8$\times$8$''$ configuration at a spatial resolution of 0.25$''$/pixel and used $J$, 
$H$ and $K$-band gratings which result in a spectral resolution of $\lambda$/$\Delta\lambda$=4000
We used ABBA chop sequences while keeping the object inside the IFU at all times.  Typical integration 
times were 7.2\,ks (split into 600 second exposures) in $<$0.6$''$ seeing and photometric conditions.  
Individual exposures were reduced using the SINFONI {\sc esorex} data reduction pipeline 
and custom {\sc idl} routines which, together, extracts, flatfields, wavelength calibrates the data 
and forms the data cube.  The final data cube was generated by aligning the individual data-cubes 
and then combining the using an average with a 3$\sigma$ clip to reject cosmic rays.  For flux calibration, 
standard stars were observed each night during either immediately before or after the science exposures.  
These were reduced in an identical manner to the science observations.  Detailed analysis of the 
spatially resolved properties will be discussed in a forthcoming paper (Livermore et al. 2011 in preparation), 
but here we concentrate on the galaxy integrated emission line properties measured from these 
observations.

\subsubsection{Line fluxes and line widths}
\label{seclines}

\begin{figure*}
\include{summary2}
\caption{\label{exspec}Typical extracted NIRSPEC  spectra (flux units in erg s$^{-1}$ cm$^{-2}$ \AA$^{-1}$) 
of various signal-to-noise ratios. For RXJ1053, the key nebular lines were covered using 3 different 
spectrograph settings (shown). Note the detection of [S{\sc ii}] emission for MACS0744. The dotted line 
indicates the 1$\sigma$ error on the spectral flux.}
\end{figure*}

Typical reduced spectra are shown in Figure \ref{exspec}. Line fluxes were measured using 
the IRAF task {\tt splot} which uses both the science spectrum and the 1$\sigma$ error spectrum 
obtained from the extraction to derive the total flux and a bootstrap error. The line widths were 
measured on the highest signal-to-noise ($>10$) spectra, and corrected for the effects of 
instrumental resolution. Results are given in Table~\ref{masses}. We fit a single Gaussian to 
the H$\alpha$ and [O{\sc iii}]$\lambda5007$ lines, and a double Gaussian to [O{\sc ii}] with the lines 
fixed at rest wavelengths of 3726.1,3728.8 \AA. We constrain the [O{\sc ii}] lines to have the same 
width and an intensity ratio $I(3726.1)/I(3728.8)\sim1$ as seen in high redshift galaxies observed with higher 
spectral resolution (e.g. \citealt{Swinbank09}).

The spectral resolution $R = \lambda/FWHM$ measured from bright, unblended OH sky lines in 
the NIRSPEC spectra is found to vary linearly with spectral order $m$ as expected, and the ratio 
$R/m$ varies smoothly with wavelength from roughly 2100 at $\lambda = 1\,\mu$m ($m=4$) to 
4000 at $2.2\,\mu$m ($m=2$). This results in an instrumental FWHM ranging from $160$--260 km/s 
($\sigma = 70$--110 km/s) for the lines observed in this work. Measured line widths exceed the 
instrumental resolution in all but one case (RXJ1347-11), for which the 1$\sigma$ upper bound 
is given in Table~\ref{masses}.

The uncertainty in the line widths is propagated from the 1$\sigma$ error spectrum. Gaussian fits 
used to determine the line width have residuals $\chi^2_{\nu} \simeq 1$ suggesting that the error 
spectrum is a good estimate of the intrinsic noise. After subtracting the instrumental resolution in 
quadrature, line widths are typically determined to $\sim$10\% accuracy. The uncertainty is significantly 
larger in cases where the line of interest is blended with a sky line, or has a width close to the 
instrumental resolution. We estimate any systematic uncertainty in the measured instrumental 
resolution to be $< 3$\% based on independent measurements of multiple bright sky lines, hence 
we expect measurement errors to dominate. 

\begin{figure}
\includegraphics[width=8.5cm,angle=0]{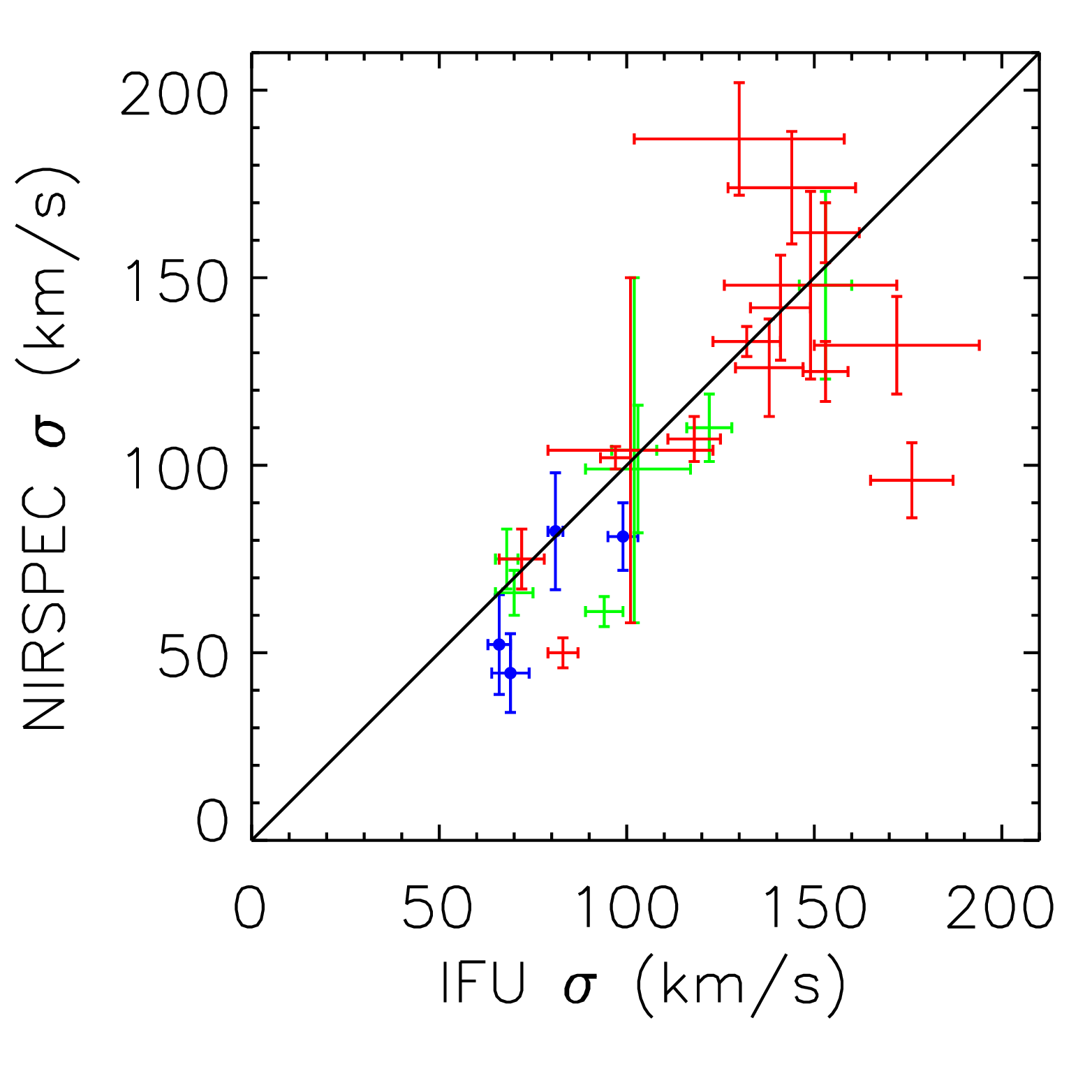}
\caption{\label{dispersion} Comparison of the velocity dispersion measured from our long slit NIRSPEC data with 
2D integral field unit data from various sources. Blue points refer to mean dispersions ($\sigma$) from OSIRIS IFU 
observations reported in \citet{Jones}. Green crosses show equivalent OSIRIS data from \citet{Law09}  
and red crosses are integrated velocity dispersions from the SINFONI IFU \citep{FS09}.
NIRSPEC measurements for the latter samples are reported in \citet{Erb06}.}
\end{figure}

We can test whether the velocity dispersions measured with NIRSPEC are reliable, where
there is overlap with integral field data, by undertaking a comparison with the more extensive 2D velocity
data \citep{Law09,FS09,Jones}. Here we enlarge the comparison
sample by considering all relevant NIRSPEC data \citep{Erb06}. As Fig. \ref{dispersion} shows, the data 
is in general agreement with the mean IFU velocity dispersion being $1.2 \pm 0.3$ times that measured 
with NIRSPEC. This suggests that the longslit data provide a reasonable estimate of the global galaxy dynamics and 
cab be used, for example, with the spatial extent to measure dynamical masses. 

\subsubsection{Line ratios from stacked spectra}

\label{composite}

One of our aims is to measure the mass-metallicity relation in our sample, extending it to low-luminosity sources. 
Given the short exposure times for our survey which has enabled such a large sample to be constructed,
inevitably many of the fainter diagnostic lines, e.g. [N{\sc ii}] and [S{\sc ii}] are not always visible in individual
spectra. In order to estimate the typical prevalence of such faint emission lines, we construct a stacked 
spectrum about the H$\alpha$ line. 

\begin{figure}
\includegraphics[width=8.5cm,angle=0]{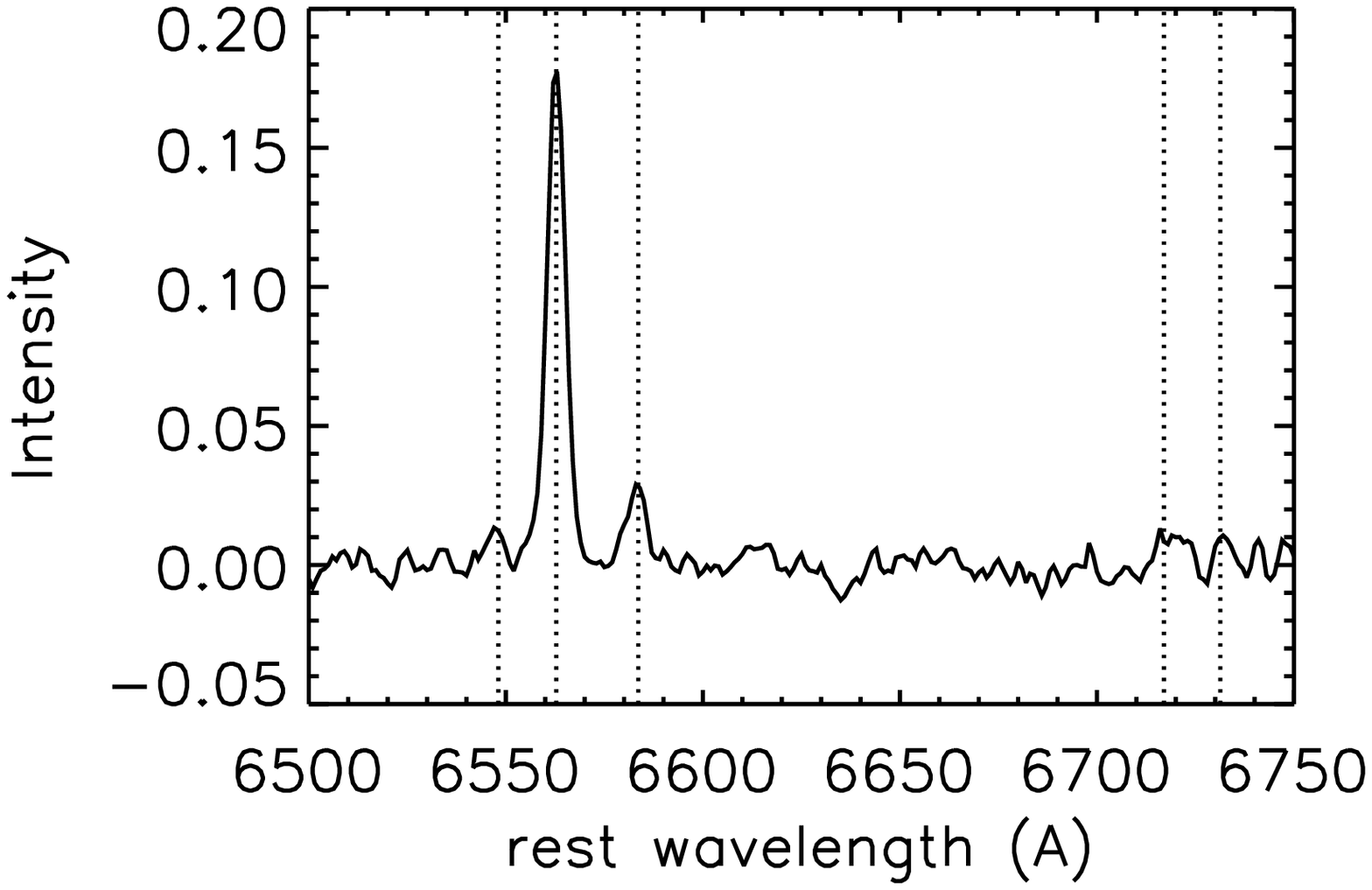}
\includegraphics[width=8.5cm,angle=0]{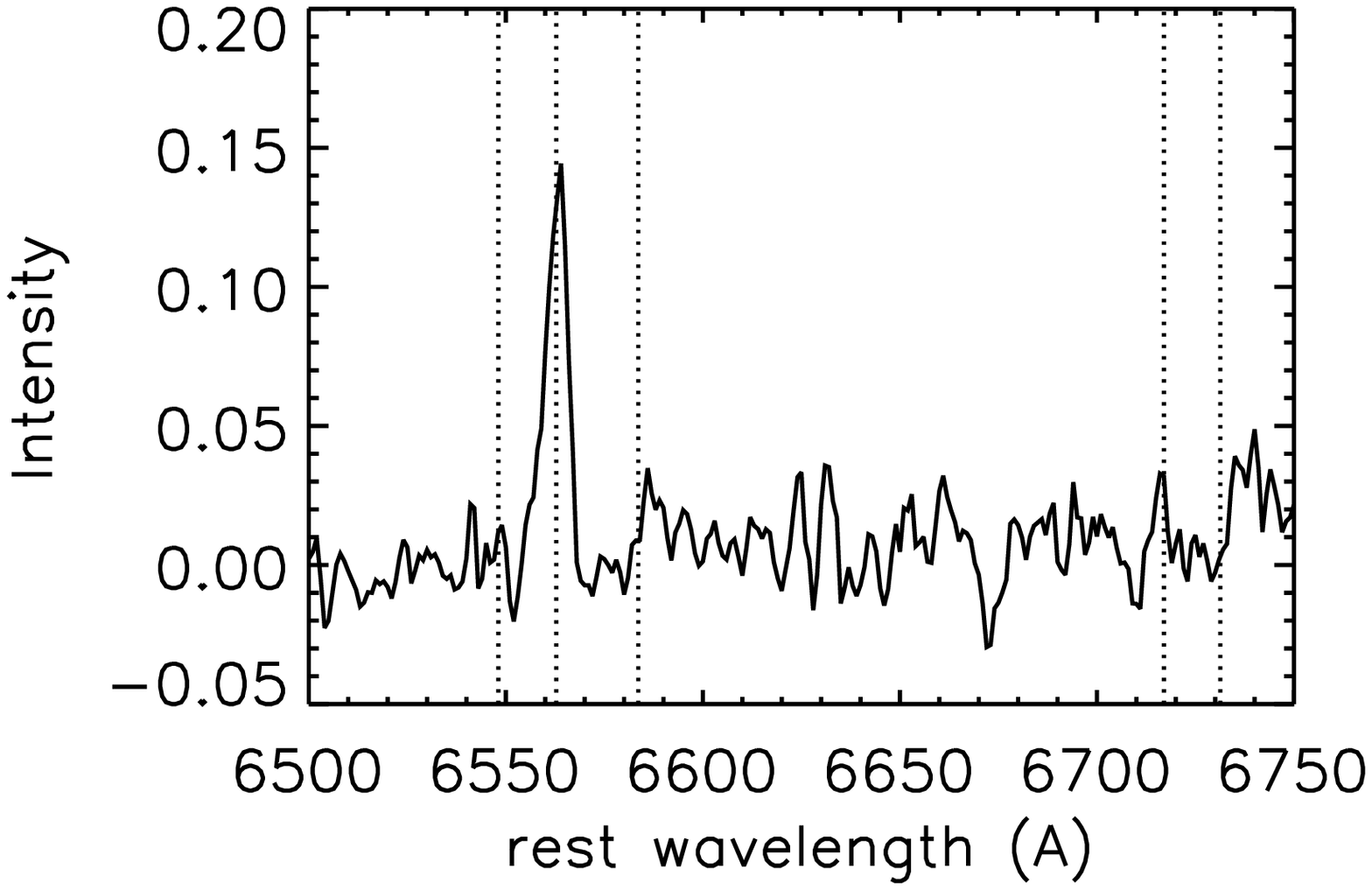}
\caption{\label{stack} Stacked spectra chosen to reveal fainter emission lines. Top: stack of all lensed sources 
with S/N $>10$ in the H$\alpha$ line. Bottom: stack of lensed sources with stellar mass $< 3\times10^9$\,M$_{\odot}$. 
Dotted lines show the positions of H$\alpha$, [N{\sc ii}]$\lambda\lambda6548,6584$, and [S{\sc ii}]$\lambda\lambda6717,6731$.}
\end{figure}

This has been done for two cases. Firstly, for all 9 lensed sources where H$\alpha$ is detected at S/N $>10$, 
and, secondly, for all 6 cases where the stellar mass (\S2.4) is $<3\times10^9$ M$_{\odot}$. First we resample 
the H$\alpha$ spectra into the rest frame with a common dispersion of $1$\,\AA, and then scale each spectra to 
a common H$\alpha$ flux. We reject the minimum and maximum values at each wavelength in order to remove 
outliers (due to sky line residuals, for example), and average the remaining data. The results are consistent with 
Gaussian noise. The composite spectra are shown in Figure~\ref{stack}.

We measure line ratios in the stacked spectra by fitting a Gaussian profile to H$\alpha$, then using the position 
and line width to determine the [N{\sc ii}] and [S{\sc ii}] fluxes and their bootstrap error. The resulting ratios  are 
provided at the bottom of Table \ref{fluxes}. 

\subsubsection{Aperture corrections and further checks}

One obvious limitation of long-slit spectroscopy is a significant fraction of the line flux may be missing, 
especially for the most extended objects or in poor seeing conditions. Our slit width was 0.76 arcsec
and so the latter may be a concern for some objects (see Table \ref{runs}). In order to compare measurements 
based on long-slit spectroscopy with photometric estimates derived from imaging (see Sect. \ref{photom}), 
we carefully estimated aperture correction factors. 

The procedure adopted was the following: we used the SExtractor \textit{segmentation map} to select those
pixels of the HST image associated with the object. We smoothed this new image by our estimate of the seeing 
conditions (Table \ref{runs}) and measured the fraction of the total flux falling inside the region covered by 
the long slit. We found aperture correction factors between 1.2 and 2.9 depending on the object. The error on these 
corrections was estimated using a $\pm0.2$\arcsec error on the location of the slit, as checked from the SCAM images. 
The correction factors are later used to derive the total star formation rate (SFR) of the object, assuming the measured 
equivalent width of the lines are representative of its average across the entire galaxy. Of course, the metallicity 
and line ratio measurements are unaffected as they rely entirely on the spectroscopic data. We also checked that measurements 
of the 8 o'clock arc are consistent with the results presented by \citet{Finkelstein}.

Six sources in our sample have been presented by \citet{Jones} and were therefore covered both with NIRSPEC 
and OSIRIS. This allows us to perform a further check on the aperture correction factors. We found that the total line 
fluxes agree for each object within 10\%, which is likely due to the aperture correction and/or the relative flux 
calibrations. We adopt this 10\% error estimate on the total line flux when estimating the SFRs later in the paper. 

As mentioned earlier, we can use the same six targets to compare the velocity dispersions estimated from the NIRSPEC 
and OSIRIS data on the same emission lines (\S2.2.3). Although the long slit estimates have larger error bars, they are consistent 
(within 1$\sigma$) with the more reliable IFU estimates. Nonetheless, we adopt the IFU value whenever availble.

\subsection{Magnification and source reconstruction}
\label{seclens}

In order to correct all physical properties (star-formation rates, masses, physical scales) for the lensing magnification,
we correct each source for its corresponding \textit{magnification factor} $\mu$ which stretches the physical scales of 
each object while keeping its surface brightness fixed. As discussed in \S2.1, our targets were specifically chosen to
lie in the fields of clusters whose mass models are well-constrained from associated spectroscopy of many lensed
sources and multiple-images (see example in Appendix A and \citet{Richard10b} for further discussion).

The values of $\mu$ are obtained through modeling of the cluster mass distribution using the {\rm Lenstool} software
\footnote{\rm http://www.oamp.fr/cosmology/lenstool/} (see references in Table \ref{targets}), or from the literature 
(for the additional targets in Table \ref{targets}). The relative error on $\mu$, 
when derived from a {\rm Lenstool} parametric model, follows a Bayesian MCMC sampler \citep{Jullo} which 
analyses a family of models fitting the constraints on the multiple images. In Table \ref{targets} we report 
the final values of $\mu$ and their associated errors; these are used to correct $\mu$-dependent physical 
parameters (star-formation rates and masses) throughout the rest of the paper. Note that the magnification factor 
does not affect any of the line ratios measurements.  
 
We also construct the demagnified (unlensed) source morphology of each arc modeled with 
{\rm Lenstool} by ray-tracing the high-resolution HST image back to the source plane, using the best 
fit lens model. By comparing the sizes of the observed and reconstructed image of a given target, we can 
verify the agreement of the magnification factor used and this size ratio 
(c.f. discussion in \citealt{Jones}). 

\subsection{Photometric measurements}
\label{photom}

A large variety of space- and ground-based images are available for each object from archival sources. 
All HST images (optical and near-infrared) have been reduced using the {\tt multidrizzle} package \citep{Koekemoer}, as well 
as specific IRAF scripts for NICMOS data, as described in \citet{Richard08}. Ground-based near-infrared 
images were reduced following the full reduction procedure described in \citet{Richard06}, and calibrated 
using 2MASS stars identified in the field.

Archival IRAC data in the first 2 channels (3.6 and 4.5 $\mu$m) are available for all sources except 
MACS1423, but we only consider sources which are not contaminated by nearby bright galaxies when 
deriving their photometry. We combine the post-BCD (Basic Calibrated Data) frames 
resampled to a pixel scale of 0.6\arcsec.

\begin{figure*}
\includegraphics[width=8.5cm,angle=270]{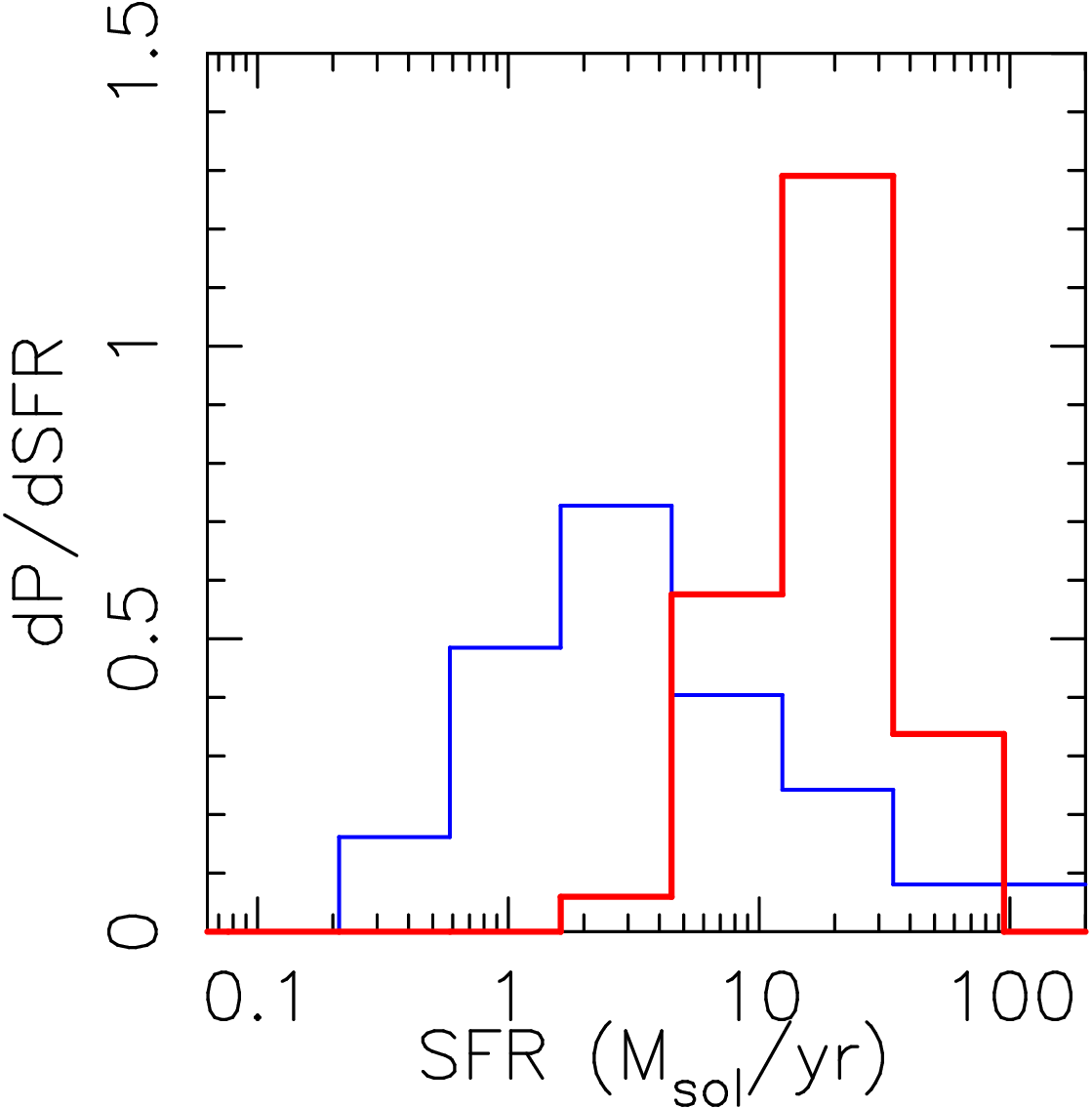}\hspace{0.2cm}
\includegraphics[width=8.5cm,angle=270]{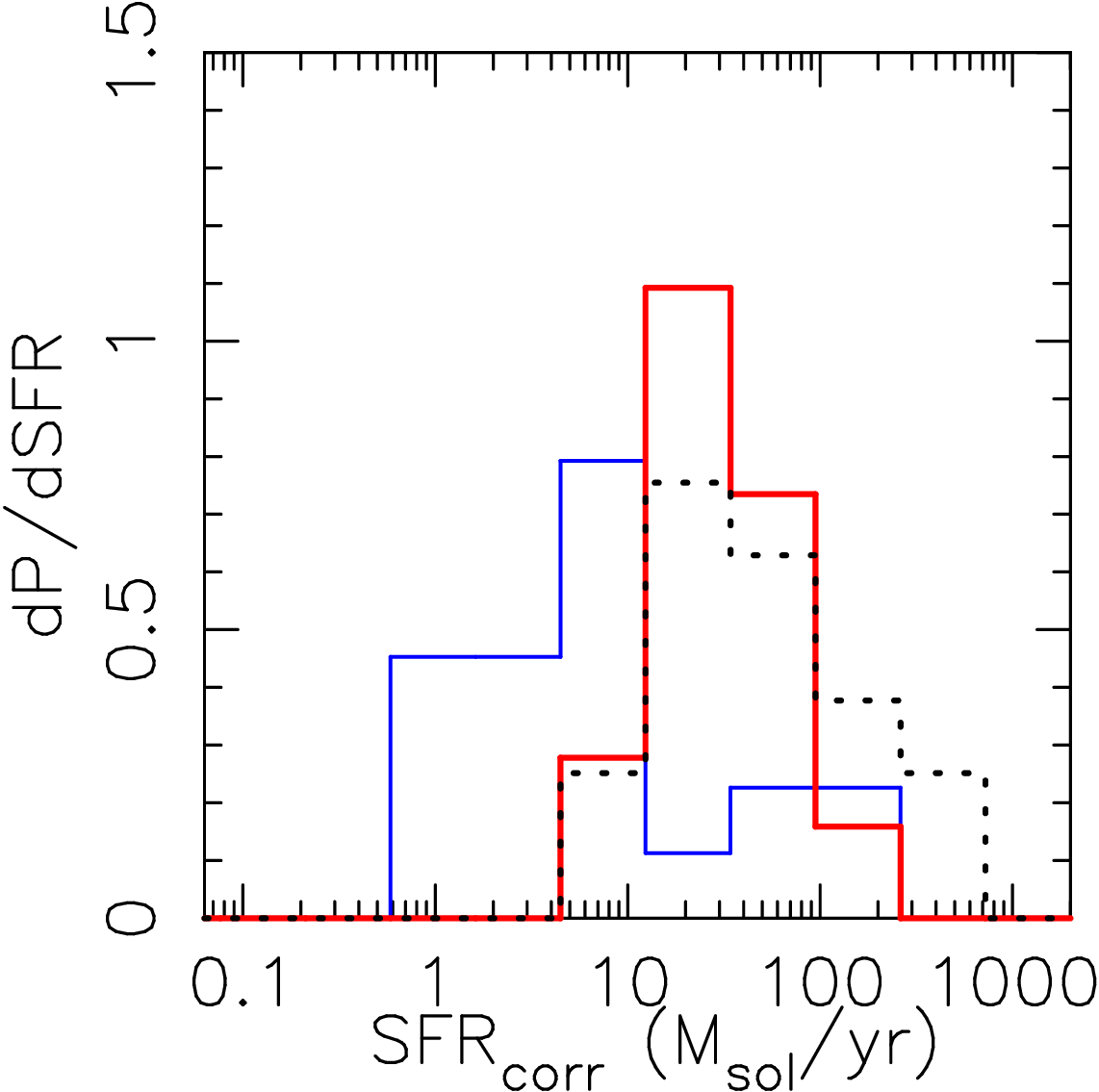}
\caption{\label{sfrs}Probability distribution of star-formation rates in our sample of lensed objects (blue histogram) 
compared to the samples  of LBGs at $z\sim2$ (red histogram, \citealt{Erb06}) and at $z\sim3$ (dotted 
histogram, AMAZE and LSD samples,\citealt{Maiolino} and \citealt{Mannucci}). The left panel 
compares the distribution of SFR before correcting for extinction, while the right panel compares the values 
after applying the extinction correction. 
}
\end{figure*}

The HST image providing the highest signal-to-noise was used to measure the integrated brightness 
with SExtractor \citep{SExtractor}, whereas the {\it double image} mode was used to measure the relative 
HST colours inside a 1\arcsec aperture. A small aperture correction is applied to all HST photometry to 
deal with the PSF differences between the different ACS, WFPC2 and NICMOS bands. In the case of 
ground-based colours, the primary HST image was smoothed by a Gaussian kernel corresponding to 
the measured seeing, and convolved with the IRAC PSF (derived from bright unsaturated stars in the image) 
in order to incorporate IRAC colours. The full photometry is summarised in Table \ref{SEDtable} in appendix.

\section{Physical Properties of the Sample}

Table \ref{masses} summarises the derived quantities which will form the basis of our analysis. We now 
discuss the physical measures in turn. 

\subsection{Star formation rate and AGN contribution}

The intrinsic star formation rate (SFR) is estimated from the total (aperture-corrected) flux in the Balmer 
lines ($f_{\rm H\alpha}$ and $f_{\rm H\beta}$) based on the well-constrained calibrations by 
\citet{Kennicutt}, including the correction for magnification factor $\mu$ and its associated error. In the absence of H$\alpha$ we 
assume a typical ratio H$\alpha/$H$\beta$=2.86. For two objects (RXJ1347 and A1689-Highz), 
neither H$\alpha$ nor H$\beta$ is available, therefore we use the [O{\sc ii}] line to derive the SFR, 
although it is a less robust estimator because of metallicity and excitation effects (e.g. \citealt{Gallagher}).

The derived values of SFR span a wide range, between 0.4 and 50 M$_\odot$/yr, and are typically a factor 
of 10 lower than other samples of LBGs with emission line measurements (Fig. \ref{sfrs}). The majority (16 objects) 
of the lensed sources have SFR$<4$M$_{\odot}$/yr, values which are absent from the samples of \citet{Erb06} and 
\citet{Maiolino}.

An independent estimate of the SFR can be obtained for every object using the ultraviolet 
continuum luminosity ($L_\nu$) estimated at 1500 \AA\ rest-frame. We measure $L_\nu$ by fitting a 
power-law $f_\lambda\propto\lambda^{-\beta}$ to the broad-band photometry 
between 1500 and 4000 \AA\ rest-frame. The UV slope ($\beta$) is given in the first column of Table 
\ref{SEDtable}. A value $\beta=2.0$ (constant AB magnitude) is assumed when 
only a single photometric datapoint was available. The SFR derived from the UV-continuum is 
then given by \citet{Kennicutt}.

\begin{figure}
\includegraphics[width=8.5cm,angle=270]{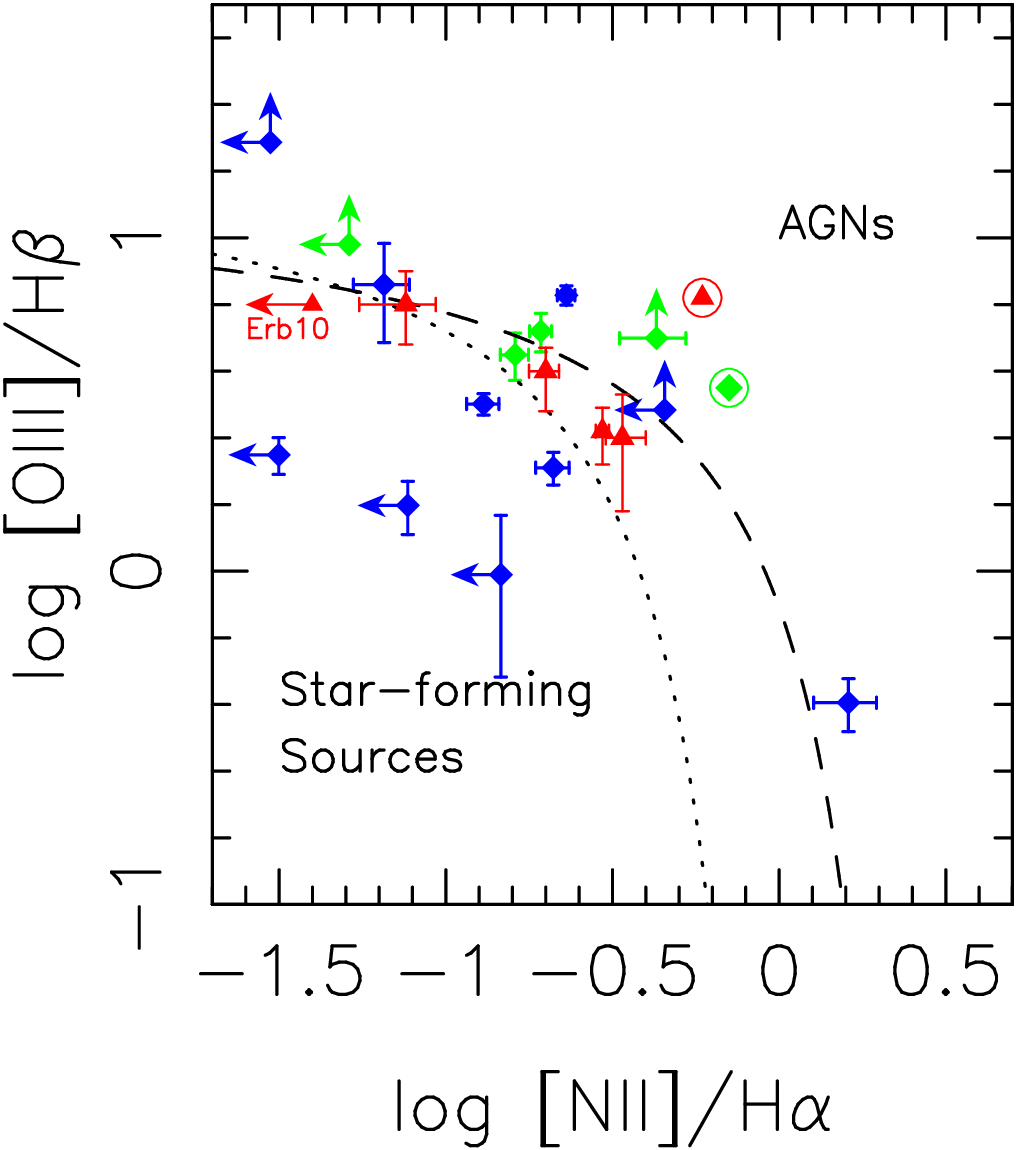}
\caption{\label{bpt} Location of the NIRSPEC targets (blue diamonds) over the BPT diagram, used as 
a diagnostic for AGN vs star-formation (see text for details). The dotted line is the empirical 
separation between star-forming and AGN objects from \citet{Kauffmann}, and the dashed line 
the theoretical separation from \citet{Kewley01}. Other sources from the literature are shown as 
green points (lensed objects from \citealt{Hainline} and {lemoine03}) or red triangles \citep{Erb06,Erb2010}. 
The two circled points are objects showing signs of AGN activity, as discussed within these papers. 
}
\end{figure}

We can determine from nebular line ratios whether there is a 
strong contribution from active galaxy nuclei (AGN) in the line emissions. This is 
done through the diagram of \citet{bpt} (BPT diagram), where we overplot the location 
of our targets in Fig. \ref{bpt}. 

The majority of our sources lie in the region of the BPT diagram where star-forming galaxies 
are commonly found \citep{Kewley01}. MACS0451 and AC114-A2, however, lie close to the boundary with 
the region occupied by AGNs. We consider it unlikely, however, that these objects are 
AGN-dominated, because of the lack of X-ray emission in Chandra data and C$_{\sc IV}$ lines in the optical spectra.
Furthermore, the Spitzer data shows evidence for the 1.6$\mu$m rest-frame stellar bump and no sign of obscured AGN activity, which would
produce a raising slope in the redder IRAC channels (e.g., \citealt{HainlineL}).
The location of these objects in the BPT plane simply suggests a higher radiation field, similar to 
that seen in local starbursts and the sample of LBGs studied by \citet{Erb06a} at $z\sim2$ 
(see also Sect \ref{ion}). We thus conclude that the nebular emission we see in our sources
arises from intense star formation.

\subsection{Extinction}

Dust extinction plays an important role when deriving the physical properties of galaxies, as it 
will affect the observed line fluxes and some of the line ratios. One of the estimators we can use to 
measure this extinction is the UV spectral slope $\beta$, which is related to the extinction affecting the young stars. 
We can also use the ratio between the two SFR estimates (from the observed UV continuum and the 
H$\alpha$ emission lines) as an independent estimator, as it reflects the differential extinction between 
the two wavelengths. Note that even in the case of a perfect reddening correction, the nebular emission 
and the UV do not probe star-formation on the same timescales \citep{Kennicutt}.

\begin{figure}
\includegraphics[width=7.5cm,angle=270]{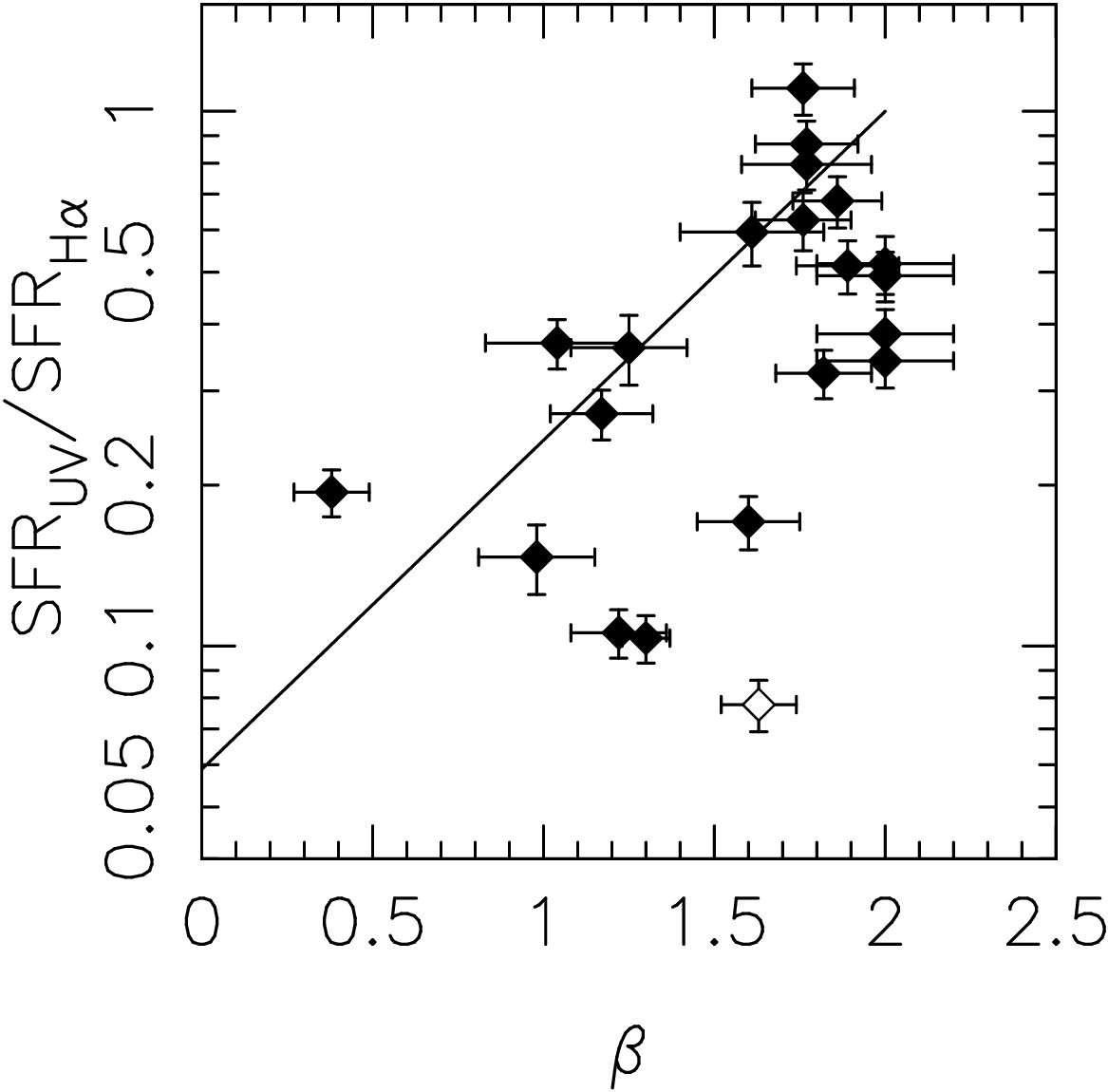}
\caption{\label{extinc}Comparison between the SFR estimated from the UV continuum and the H$\alpha$ emission 
line, as a function of the UV slope $\beta$. The solid curve is the theoretical prediction for the relation 
between the two values using the \citet{Calzetti} extinction law.}
\end{figure}

Figure \ref{extinc} compares both estimates for a subsample of the lensed objects, together with the relation 
predicted by the \citet{Calzetti} extinction law. It shows that there is a general agreement with the theoretical 
predictions, although with quite a large scatter. One of the reasons for the differences is probably a 
different extinction factor affecting the young and old stellar populations, or a measurement bias towards 
low extinction regions when measuring $\beta$. This is illustrated by the location of the sub-millimetre 
lensed galaxy A2218-smm \citep{Kneib04b}, which has a high extinction but a measured slope $\beta=1.6$ 
(white diamond in Fig. \ref{extinc}).

One way to overcome this issue is to use the full SED (from rest-frame UV to near-infrared) in order to derive the best 
extinction estimate E(B-V) assuming the \citet{Calzetti} law (see Sect.\ref{sedfit}). 8 objects in our sample also have both H$_\alpha$ 
and H$_\beta$ detected, and the corresponding values of the Balmer decrement range between 2.5 and 4.5, 
consistent with $0<E(B-V)<0.35$ although with large uncertainties. 

We use the best-fit E(B-V) values, given in Table \ref{masses}, 
to correct the SFRs, individual line ratios and metallicities in the next sections. In particular, reddening has a strong effect 
on individual SFRs, and it is very sensitive to unknown factors (differential reddening in a given object, presence 
of dust in the lightpath through the galaxy cluster, deviation from the Calzetti law as pointed out by \citealt{Siana}). 
We present both the uncorrected and the extinction-corrected (SFR$_{corr}$) in Table \ref{masses} and when comparing 
with other samples (Fig. \ref{sfrs}). 

\subsection{Masses}

Two mass estimators can be derived from the available data. Multi-wavelength broad-band photometry 
gives us access to the stellar mass, through the modelling of the SED, while measuring the widths of 
the most prominent spectral lines allow us to infer dynamical masses, which should be closer to the 
total baryonic mass of these galaxies.

\subsubsection{Stellar masses}
\label{sedfit}
Our stellar masses are derived using the precepts discussed in
detail by \citet{Stark09}. 
We derive the stellar masses for our sample by fitting the Charlot \& Bruzual
(2007, S. Charlot, private communication) stellar population synthesis models to the observed SEDs.   
We consider exponentially-decaying
star formation histories  with the form $\rm{SFR(t)}\simeq exp(-t\tau)$
with e-folding times of  $\tau=10$, 70, 100, 300, and 500 Myr in addition
to models with continuous  star formation (CSF).    For a given galaxy, we
consider models ranging in age from 10 Myr to the age of the Universe at
the galaxy's redshift.   We use a \citet{Salpeter} initial mass
function (IMF) and the \citet{Calzetti} dust extinction law.   
Finally, we allow the metallicity to vary between solar (Z$_\odot$) and
0.2 Z$_\odot$, the range found for the gas-phase metallicity using nebular 
line ratios (Sect. \ref{z}). We account for the intergalactic medium absorption following \citet{Meiksin}. 
The best fit values of the stellar mass M$_*$ and extinction E(B-V) are summarised in Table 
\ref{masses}.

We note that the presence of strong emission lines may affect the broad-band photometry and therefore 
the mass estimates. We estimate the contribution of the strongest emission lines affecting the $H$ and 
$K$ band magnitudes. In the most extreme cases (largest equivalent widths) the H and K band flux 
would be affected by at most 5-10\%, which does not have a significant effect compared to our estimated errors.

\subsubsection{Dynamical masses}

We now use the velocity dispersions $\sigma$ (measured in Sect. \ref{seclines}) to  estimate 
of the dynamical mass. We will express the virial masses M$_{dyn}$ of our objects as a function 
of $\sigma$ and their typical  size. 

Assuming the idealized case of a sphere of uniform density \citep{Pettini} we have:
\medskip\par

\hspace{-0.65cm}$M_{dyn} = C r_{1/2} \sigma^2 / G$ with $C=5$ or more conveniently,
\begin{equation}
\label{dyneq}
M_{dyn}=1.16\times10^{10}\ M_\odot \frac{\sigma^{2}}{(100\ {\rm km}\ {\rm s}^{-1})^{2}}\frac{r_{1/2}}{\rm kpc}
\end{equation}

where $r_{1/2}$ is the half-light radius, which we measure on the source plane reconstructions of 
our targets (see Sect. \ref{seclens}) using the {\tt FLUX\_RADIUS} parameter from SExtractor. This parameter 
estimates a circularized size corresponding to half of the total detected fluxes. Thanks to the high magnification 
our sources are well resolved in the HST images, but the main source of error in estimating $r_{1/2}$ (which has 
a strong impact on M$_{dyn}$) is the source reconstruction itself. We produced 100 reconstructions of each 
source sampling the different parameters of the lens model, using the MCMC sampler described in Sect. \ref{seclens}, 
and use them to derive the mean and dispersion on the measured $r_{1/2}$. We also checked this measurement 
independently using a half-light radius defined with the Petrosian radius at 20\% of the central flux of the object, 
similar to the work done by \citet{Swinbanksizes}, and found consistent results.

The values of $r_{1/2}$ and $M_{dyn}$ are 
reported in Table \ref{masses}.  We note that the geometric correction factor C can vary significantly and is 
likely between 3 and 10 for these highly turbulent galaxies \citep{Erb06}. Hence the true dynamical mass can differ by up 
to a factor of 2, while uncertainty in the radius and $\sigma$ is much smaller. Our choice of $C=5$ gives slightly 
higher dynamical masses than in some other studies (e.g. $C=3.4$, \citealt{Erb06,Bouche}).

\subsubsection{Comparison}

\begin{figure*}
\includegraphics[width=8.cm,angle=270]{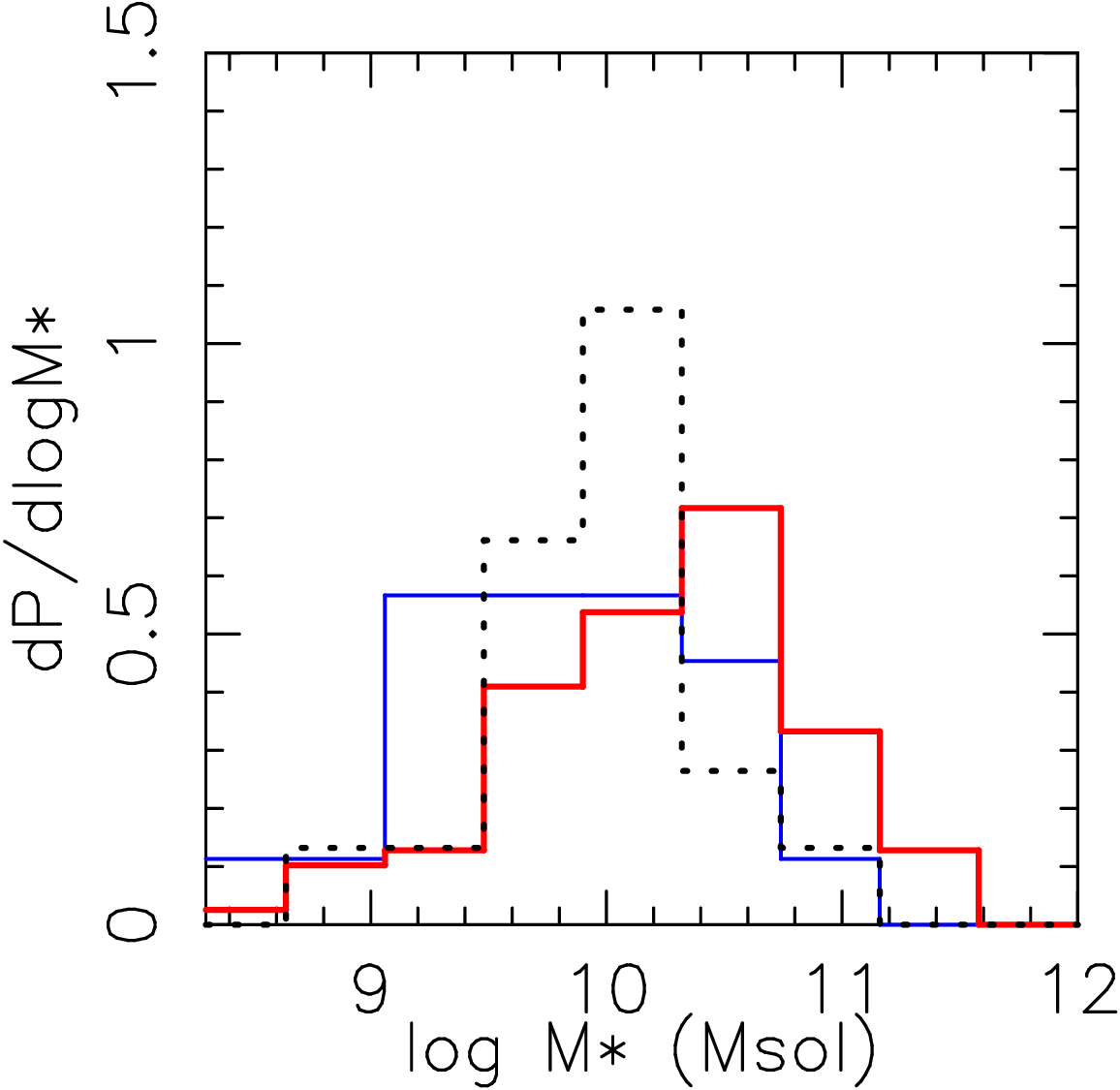}
\includegraphics[width=8.cm,angle=270]{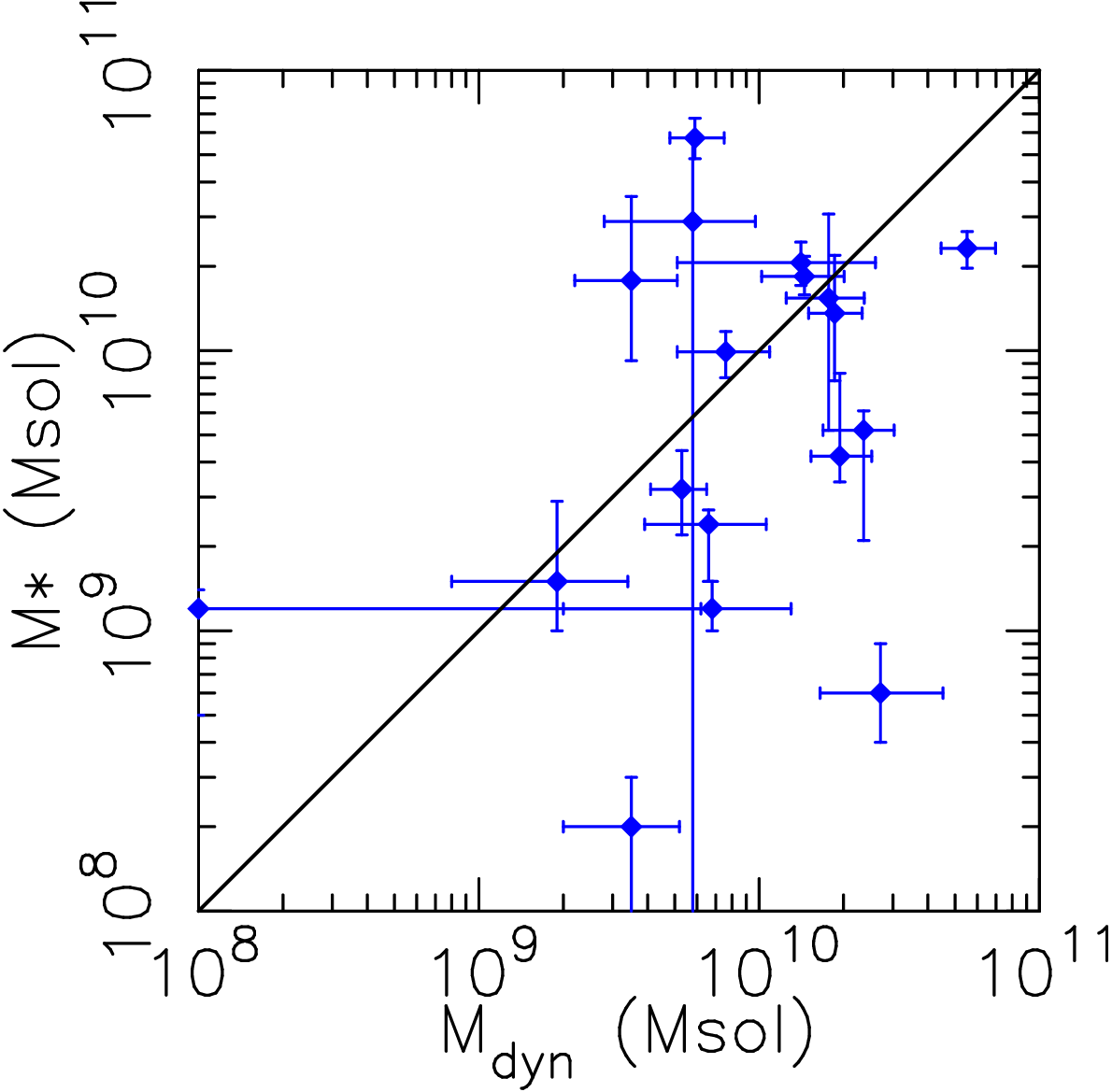}
\caption{\label{mm}
(Left) Probability distribution of stellar  masses in our sample of lensed objects (blue histogram) compared to the samples  
of LBGs at $z\sim2$ (red histogram, \citealt{Erb06b}) and at $z\sim3$ (dotted histogram, AMAZE and LSD samples, \citealt{Maiolino} and \citealt{Mannucci}).
(Right) 
Comparison between stellar and dynamical mass estimates. In general we find a higher 
dynamical mass compared to the stellar mass, suggesting the presence of large amounts of gas mass in this 
sample.
}
\end{figure*}

The range of stellar masses spanned by our lensed objects (typically 5\ 10$^8$ - 5\ 10$^{10}$ M$_\odot$ is lower 
by a factor of 3-5 than those surveyed by \citet{Erb06b,Maiolino} and \citet{Mannucci} (Fig. \ref{mm}, left). Within the subsample 
with reliable dynamical masses, we can directly compare the dynamical and stellar masses. Although we discover
a large dispersion (Fig. \ref{mm}, right), the average ratio  $<log(M_{dyn}/M*)>=0.36$ dex is a factor of $\sim2$, 
suggesting of $\sim$40\% of dynamical mass not present into stars. This value is similar to the result found by 
\citet{Erb06b}, and confirms the dominance of baryonic mass in the central regions of these galaxies. This was 
already pointed out by \citet{Stark08} in the case of the Cosmic Eye. We note that a different choice for the multiplicative 
factor in Eq. \ref{dyneq} would give $\sim30$\% lower dynamical masses and would strengthen this conclusion.

The gas mass can be evaluated from the SFR and the size of the objects $r_{1/2}$ (ideally assuming the star-formation is 
uniform over the projected surface seen in the UV), assuming the \citet{Kennicutt} law between star-formation rate and
gas densities applies at these redshifts. We use this relation to derive the total gas mass:

\begin{equation}
M_{\rm gas}=5.03\ 10^8\ {\rm SFR}^{0.71}\ r_{\rm 1/2}^{0.58}\ {\rm M}_{\odot}
\end{equation}

Defining the gas fraction $f_{\rm gas}$ as $f_{\rm gas}=M_{\rm gas}/(M_{\rm gas}+M_*)$ , we find 
gas fractions ranging from 0.2 to 0.6 after reddening correction, compatible with what is derived from 
the comparison of stellar and dynamical masses. This is in average $\sim$50\% lower that the gas fraction 
measured by \citet{Erb06b} at $z\sim2$ and by \citet{Mannucci} in the Lyman-break galaxies Stellar populations and Dynamics (LSD)
 sample at $z\sim3$, and reflects 
the lower SFRs in our objects for a given stellar mass. However, we 
see the same trend of a lower $f_{\rm gas}$ towards the higher masses.


\subsection{Metallicities}
\label{z}
The relationship between the oxygen abundance, or gas metallicity $Z$ defined as 
log(Z)=12+log(O/H), has been accurately calibrated against  line ratios of the 
prominent nebular emission lines \citep{Nagao}. Suitable line ratios are available for 
half of the objects in our sample, and depending on the availability we can combine the 
estimates from three different metallicity diagnostics, using the recent empirical relations 
derived  by \citet{Maiolino} (see also  \citealt{Mannucci}).

The prime oxygen abundance indicator is the R23 ratio \citep{Pagel}: 
R23=([O{\sc ii}]+[O{\sc iii}]$\lambda\lambda$4959,5007)/H$\beta$. As the [O{\sc iii}]$\lambda$4959 
line is usually weakly detected we adopt a canonical value of 0.28 for the  [O{\sc iii}]$\lambda$4959/[O{\sc iii}]$\lambda$5007 
ratio. The second estimate for oxygen abundance is the O32 ratio: O32=[O{\sc iii}]$\lambda$5007/[O{\sc ii}].
The final estimate is the N2=[N{\sc ii}]/H$\alpha$ indicator.

Defining the parameter $x=log(Z/Z_{\odot})$ and using $log(Z_\odot)$=8.69 \citep{Allende}, 
we adopt the following fitting formula \citep{Maiolino} on the reddening corrected line ratios:

\hspace{-0.6cm}$\log$({\rm R23})=
\begin{equation}
0.7462-0.7149\ x-0.9401\ x^2-0.6154\ x^3-0.2524\ x^4
\end{equation}
\begin{equation}
\log({\rm O32})=-0.2839-1.3881\ x-0.3172\ x^2\\
\end{equation}
$\log$({\rm N2})=
\begin{equation}
-0.7732+1.2357\ x-0.2811\ x^2-0.7201\ x^3-0.3330\ x^4
\end{equation}

These equations provide three different estimates for Z: $Z_{\tt R23}$, $Z_{\tt O32}$ and $Z_{\tt N2}$.
Of these, the $Z_{\tt N2}$ and $Z_{\tt R23}$ estimates are the ones showing the lowest 
dispersion (typically 0.1 dex) while the $Z_{\tt O32}$ estimate has a dispersion of 0.2-0.3 dex. However, the 
$Z_{\tt R23}$ relation has a ``two-branch degeneracy" (e.g. \citealt{Pettini}) between a low-metallicity 
and a high-metallicity value. In order to discriminate between the two values, we use the $Z_{\tt N2}$ calibration 
when available (usually synonym of a source on the upper-metallicity branch), otherwise the $Z_{\tt O32}$ 
calibration. For the 5 sources taken from the literature, we recalculate the best fit Z using the published line fluxes 
and reddening factors. The resulting value of Z is consistent with the published value except in the 
case of AC114-S2 \citet{Lemoine03}, where the published value $log(Z)=7.25\pm0.2$ makes this object a clear outlier in the mass-metallicity 
relation (Sect. \ref{secmassz}). Taking the published line fluxes, the N2 limit derived places it at the junction between 
the two branches of the R23 relation. To resolve this ambiguity we use the values derived with the upper-branch.

The final best-fit value for $\log(Z)$ is reported in Table \ref{masses} for each source.The error bars include the typical 
dispersion in the relation. We find metallicity values ranging from 0.25 Z$_\odot$ to 1.7 Z$_\odot$. 

\subsection{Ionization parameter}
\label{ion}

In the previous section, the O32 parameter has been used mainly to discriminate between the lower and upper-branch of the 
R23 calibration of oxygen abundance. For a given metallicity, the O32 line ratio can also be compared with models of H{\sc ii} 
regions to measure the ionization parameter $U$, i.e. the ratio of density of ionizing photons over the density of hydrogen atoms 
\citep{KewleyDopita}. Based on early samples of bright LBGs from \citet{Pettini}, high ionization levels have been measured 
by \citet{Brinchmann08}, with $\log(U)\sim-2.0$, compared to local samples. This same ionization parameter shifts the objects upward 
in the BPT diagram (Fig. \ref{bpt}), perhaps pushing the frontier between star-forming galaxies and AGNs \citep{Erb06}.

The O32 parameter can be measured for 6 objects in our sample, as well as 2 upper limits, and  
 we obtain values of O32 in the range $0<$O32$<1$. For the range of metallicities 
found previously (0.2 to 2.0 Z$_\odot$), we use the curves provided by \citet{KewleyDopita} to derive ionization parameters 
$-2.9<\log(U)<-2$ whereas the typical values from local galaxies are in the range $-4<\log(U)<-3$ \citep{Lilly}.

\begin{figure}
\includegraphics[width=8.5cm,angle=270]{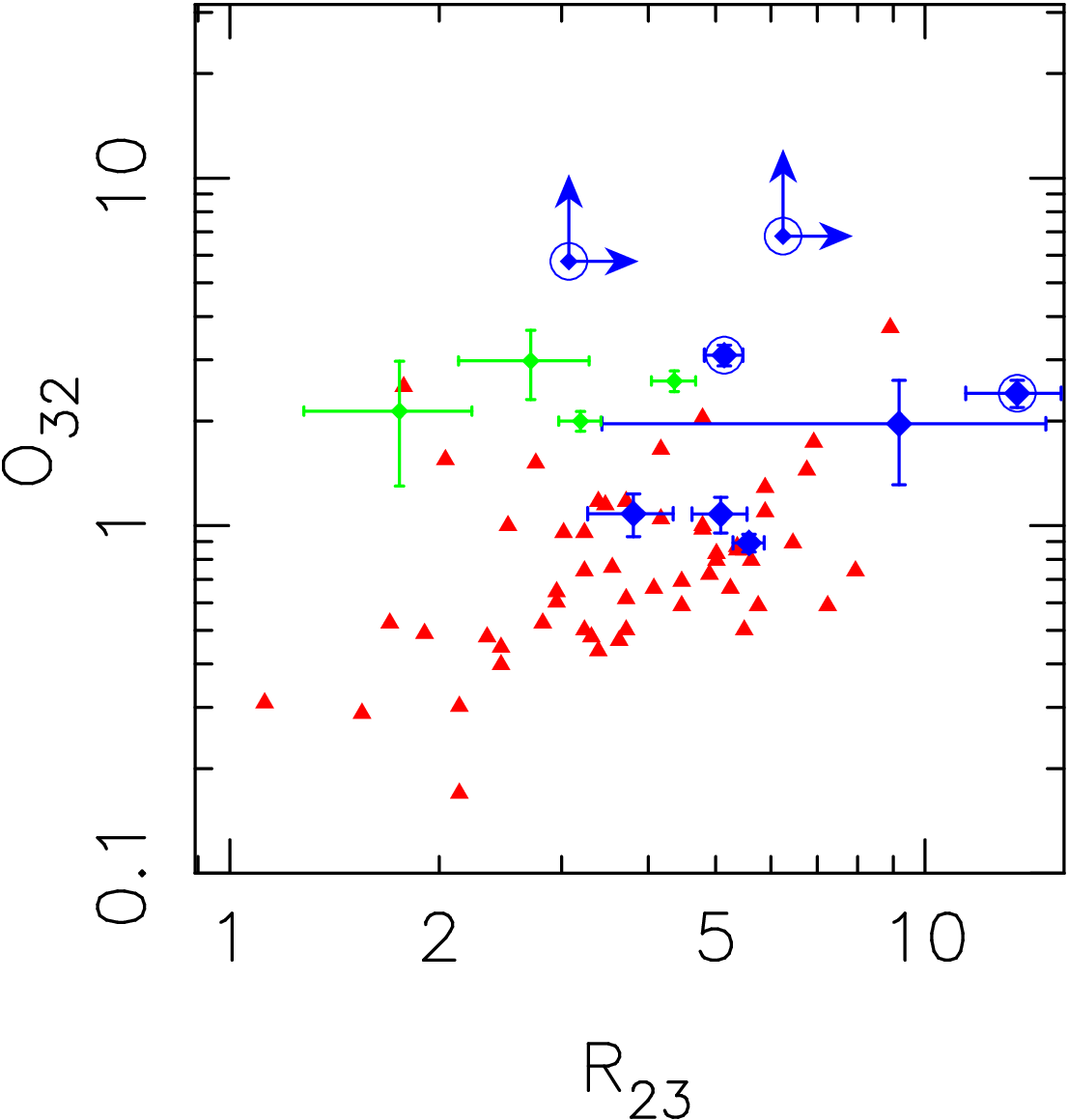}
\caption{\label{hainplot}O32 vs R23 diagram, showing the effect of the ionization parameter on O32 (adapted from \citealt{Hainline}). 
The red triangles are the low redshift ($z<1$) sample of galaxies from \citet{Lilly}. Our sample 
of $z>$1.5 lensed galaxies is shown in blue diamonds, while other literature data used by \citet{Hainline} is shown in green. 
Sources in our sample showing young stellar populations (best age $<100$ Myrs from the SED fitting) are circled and typically lie in the top part of the diagram.}
\end{figure}

This can be illustrated by constructing the O32 vs R23 diagram, as the R23 metallicity estimator has only a weak dependance 
on the ionization parameter. We overplot the results found by \citet{Hainline} on 4 lensed objects (included in our sample) and 
add our 6 new measurements and 2 upper limits in this diagram (Fig. \ref{hainplot}). We can see that , in average, our sample is systematically shifted towards higher values of O32 
compared to the lower redshift objects from \citet{Lilly}. A likely explanation for this effect might be that the physical conditions 
in the relevant H{\sc ii} regions are  different from those in the local Universe, for example, with larger electron 
density and/or larger escape fraction \citep{Brinchmann08}. 

An even more extreme result was found recently in a high redshift object 
by \citet{Erb2010}, where they derive an ionization parameter $log(U)\sim1.0$. The very young age found in this object 
is one of the factors explaining such a high value of U. Indeed, we can see some trend with age in the O32 vs R23 diagram, 
despite the small number of objects in our sample. By selecting sources having very young stellar populations (best age
$<100$ Myrs from our SED fitting), they all lie in the top part of this diagram, with the highest O32 values. 

\section{The Mass Metallicity Relation}

\subsection{Comparison with Earlier Work}
\label{secmassz}

\begin{figure*}
\includegraphics[width=8.cm,angle=270]{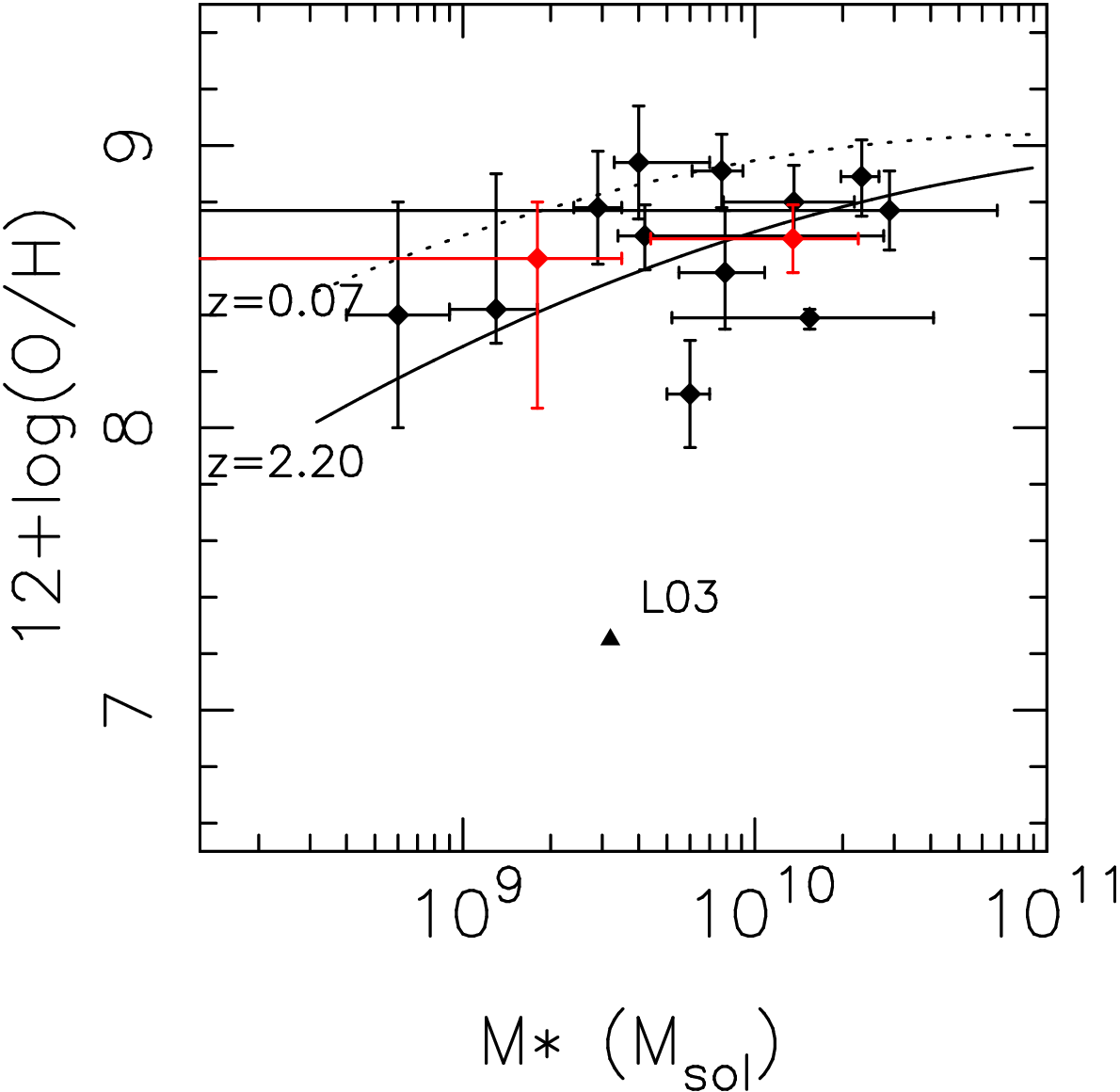}
\hspace{0.2cm}
\includegraphics[width=8.cm,angle=270]{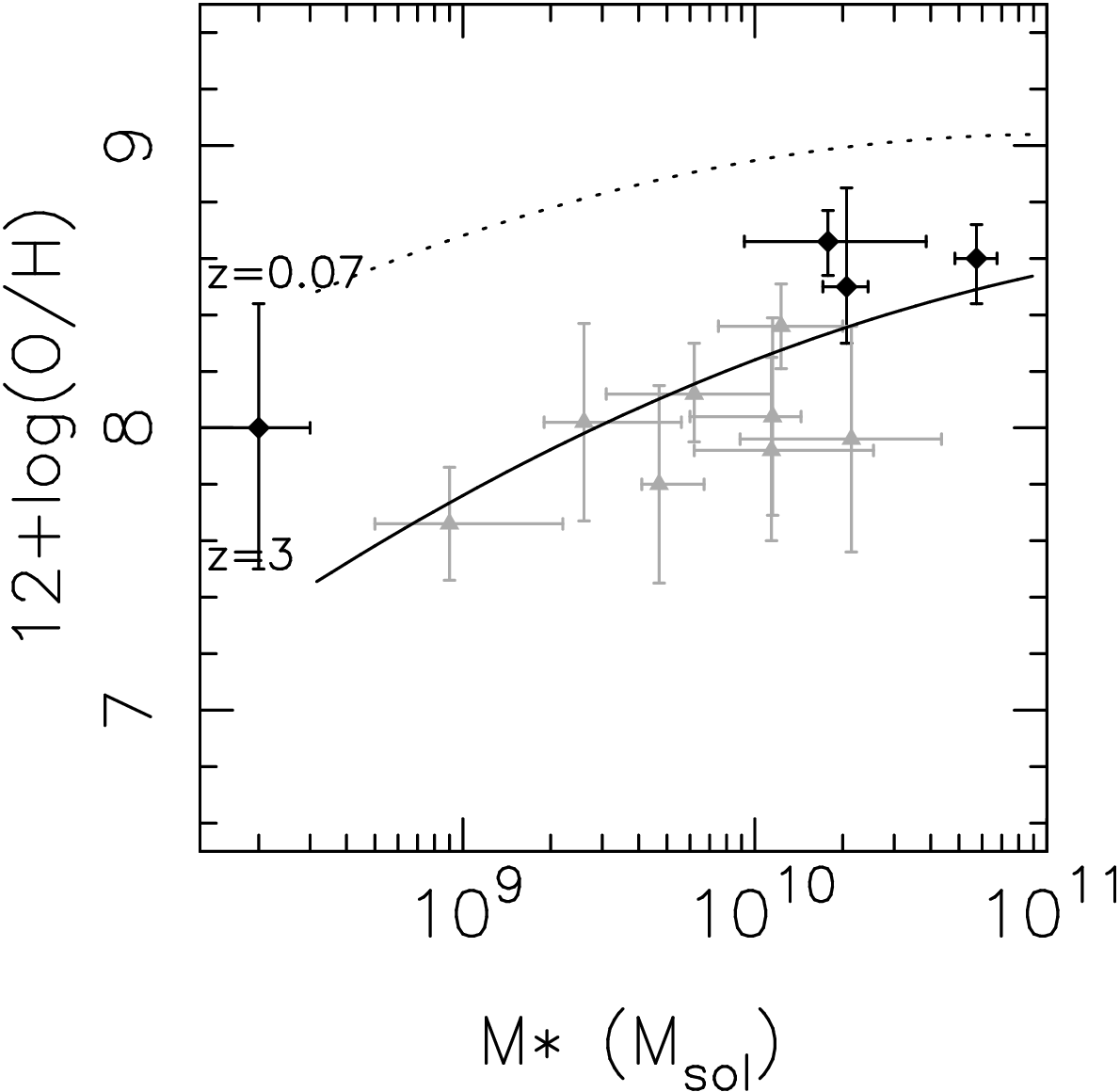}
\caption{\label{massz} The relation between gas metallicity and stellar mass as a function of redshift. This 
is based on the diagram presented by \citet{Mannucci}, with their estimates on $z\sim3$ galaxies (LSD sample) 
shown as grey datapoints. The local mass-metallicity relation is showed as a dotted line for $z=0.07$, while 
their best fit at $z\sim2.2$ and $z\sim3.0$ (including both the LSD and AMAZE samples) are shown as solid lines in 
the left and right panels respectively. We present as diamond symbols the 
subsample of $1.5<z<2.5$ lensed galaxies in the left panel, and the $2.5<z<3.5$ sample in the right panel.
The values derived from the two composite spectra (see Sect. \ref{composite}) are shown as red points in the left 
panel. The original value of $12+log(O/H)$ for AC114-S2 published by \citet{Lemoine03} is given as a triangle. }
\end{figure*}

We now compare our measurement of the mass-metallicity relation at high redshift with that of earlier workers
recognizing that our data, for the first time, probes to lower masses and lower SFRs by virtue of our selection of gravitationally-lensed
systems (Fig. \ref{sfrs} and \ref{mm}). Figure \ref{massz} summarizes the current situation. We find that all galaxies in our sample lie below the well-defined 
metallicity relation at $z=0.07$ \citep{Tremonti04} therefore supporting strongly the case for evolution. \citet{Mannucci} 
have proposed two best fits for this relation, at $z=2.20$ and $z=3.0$, from the \citet{Erb06a}, the AMAZE \citep{Maiolino} and the LSD  \citep{Mannucci} samples. 
If we partition our sample into two redshift bins: $1.5<z<2.5$ and $2.5<z<3.5$, we can see that, on average, the mass-metallicity 
relation follows these trends. However we observe a large scatter ($\sim0.25$ dex) in the $1.5<z<2.5$ range and our
the evolution implied by our data is less extreme over 2.20$<z<$3.3 than that suggested by \cite{Mannucci}.

Compared with the earlier unlensed studies, our sample spans a wider range of stellar masses. Therefore, even if 
our median stellar mass is similar to the $z\sim3$ sample from \citet{Mannucci}, we have significantly increased the 
number of galaxies at $M\sim10^9$ $M_\odot$ for the most  highly magnified objects. At these masses, we do not 
seem to see a steep decline in metallicity seen in the $z=2.2$  and $z=3$ best fit found in the earlier surveys 
(Figure \ref{massz}). Instead, we see an average shift of $\sim$0.25 dex towards higher metallicities for low-mass 
objects ($\sim10^9$ M$_\odot$). Although this trend is limited by the small statistics of our sample, 
it is also visible when overplotting the results from the composite spectra defined in Sect. \ref{composite}.

\subsection{A fundamental metallicity relation?}

By comparing the histograms in Fig. \ref{sfrs} and Fig. \ref{mm}, it is clear that our lensed objects have much lower SFRs 
than those in more luminous LBGs, even when accounting for the reddening correction, but only slightly smaller masses. Indeed, the effect of 
the increasing SFR at higher redshift is a usual explanation for the evolution of the mass-metallicity relation seen in 
Fig. \ref{massz}, and selecting objects of lower SFR would remove this effect and explain the slight deviation of our sample towards 
slightly higher metallicity. The reason for the much lower SFR in our sample is expected, as the un-lensed samples 
of LBGs are selected through their SFR in the UV, while we made no strong assumptions on this parameter. 

Recently, \citet{Mannucci2010} have proposed to include the SFR as a third component in the mass-metallicity relation, which 
would explain its evolution in redshift. One of the reason for the influence of star-formation is that \textit{outflows} would be more efficient in low-mass objects. Exploring the third-dimensional space defined by stellar mass, gas metallicity 
and SFR, they fit a surface in low redshift galaxies defined by:

\begin{equation}
{\rm Z_{est}}=8.90+0.37\ m-0.14\ s-0.19\ m^2+0.12\ m\ s-0.054\ s^2
\end{equation}

with $m=log(M*)-10$ and $s=log({\rm SFR})$. 

By comparing the measured metallicity Z with the projection Z$_{\rm est}$ estimated 
from this surface, they managed to reduce the dispersion in SDSS galaxies to $\sim0.05$ dex.   
Further computing the distance of higher redshift samples from this surface, they found no significant evolution in this 
\textit{fundamental relation} upto $z\sim2.5$, within the 1$\sigma$ error.  

\begin{figure}
\includegraphics[width=8.5cm,angle=270]{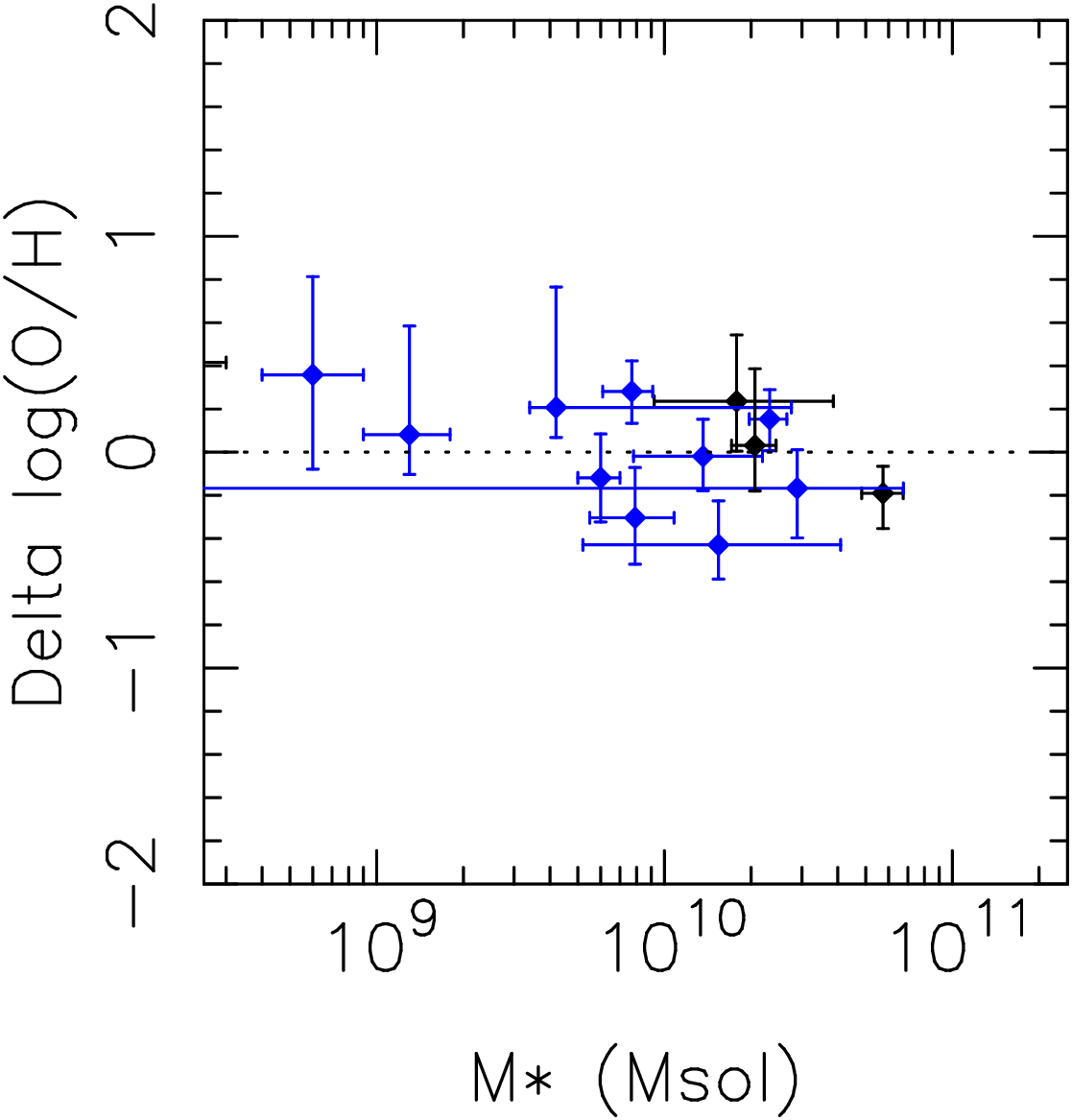}
\caption{\label{plane}Variation between the measured metallicity and the \textit{fundamental metallicity relation} 
proposed by \citet{Mannucci2010}, as a function of the stellar mass. Blue/black points are the $1.5<z<2.5$ and 
 $2.5<z<3.5$ subsamples, respectively.The fundamental relation predicts the average 
metallicity of our lensed galaxies in both redshift ranges.}
\end{figure}

We computed the values of Z$_{\rm est}$ for each galaxy in our sample and compare it with the measured Z. We find that 
in average, the best-fit surface predicts the metallicity with no significant offset, and with a scatter 
of $\sim0.2$ dex, making this metallicity relation compatible with our measurements 
within a 2$\sigma$ level. When plotting the distance to the fitted surface as a function of stellar mass (Figure 
\ref{plane}) we see no significant trend with stellar mass in both redshift ranges probed by our sample. 

\subsection{Summary and perspectives}

We have presented the results of a near-infrared spectroscopy survey targetting 23 bright lensed galaxies at $z>1.5$, 
complemented by 5 sources from the literature. We summarize here our findings: 

\begin{itemize}
\item{After correction for the magnification factor, our sample shows in average 10$\times$ smaller star-formation rates 
and 5$\times$ smaller stellar masses than the samples of LBGs at the same redshifts. Such low values of SFR would 
not be accessible without the strong lensing effect, making our sample complementary to LBG studies.}
\item{The comparison of dynamical and stellar mass estimates reveals the presence of significant gas fractions ($\sim40$\% in average), which are compatible with a simple estimation from their star-formation rate density assuming the \citet{Kennicutt} law. We observe typically lower gas fractions in the high mass objects.}
\item{We estimate the ionization parameter $U$ for 8 objects where O32 and R23 line ratios are available, and derive high 
values with $\log(U)\sim-2.5$. The highest ionization values seem to correlate with the youngest stellar 
populations ($<100$ Myrs).}
\item{The gas-phase metallicites are calculated combining various line-ratios estimators, and we find a weaker evolution in the mass-metallicity relation 
compared to estimates from bright LBGs observed in blank fields, with an offset reaching $\sim0.25$ dex in the low stellar 
mass range ($\sim10^9$ M$_\odot$). This effect is seen both in the majority of individual sources 
as well as in composite spectra created from the highest signal-to-noise or the low-mass objects.}
\item{Assuming that the evolution in the mass-metallicity relation is due to the increasing SFR at higher redshifts, we can 
reconcile our results with the existence of a fundamental relation of mass, metallicity and SFR as proposed by \citet{Mannucci2010}. 
The weaker evolution in the mass-metallicity relation in our sample is due to lower SFRs (compared to other luminous samples) 
for only slighly smaller masses.}
\end{itemize}


We foresee that the next HST programs on lensing clusters will continue to detect large number of magnified 
high redshift galaxies as multiple images, which would be ideal targets for deeper multi-object spectroscopic 
similar to the current work. This is a unique opportunity to extend the current sample to much lower stellar masses 
(typically 10$^8$ M$_\odot$) and consequently provide more constraints on the mass-metallicity relation for 
a wide range of star-formation rates, metallicities and redshifts.  

\section*{Acknowledgments}
 We acknowledge valuable comments from Filippo Mannucci which improved the content and clarity of the paper, 
and helpful discussions with Fabrice Lamareille. We are grateful to Steven Finkelstein for help in comparing our results on the 8 o'clock arc.
JR acknowledges support from an EU Marie-Curie fellowship. DPS acknowledges support from an 
STFC Postdoctoral Research Fellowship. Results are partially based on observations made with the NASA/ESA Hubble Space
Telescope, the Spitzer Space Telescope, and the Keck telescope.The authors recognize and acknowledge
the very significant cultural role and reverence that the summit of
Mauna Kea has always had within the indigenous Hawaiian community. We
are most fortunate to have the opportunity to conduct observations
from this mountain. The Dark Cosmology Centre is funded by the Danish National
Research Foundation.

\bibliography{references}

\begin{table*}
\begin{tabular}{lllclcl}
Target                       & R.A. & Dec. & $z$      & Ref. $z$ & $\mu$                            & Ref. $\mu$ \\
                                    &  (2000.0) & (2000.0) &           &           &       (mags)  \\
\hline
A68-C1                     & 00:37:06.203 & +09:09:17.43 &  1.583  &    (1)         & 2.52$\pm$0.1             &  (13)\\
CEYE                        & 21:35:12.712 & $-$01:01:43.91 &   3.074 &    (2)         & 3.69$\pm$0.12       &  (19) \\
8OCLOCK                & 00:22:41.009 & +14:31:13.81 &   2.736  &    (3)         & 2.72$^{+0.8}_{-0.4}$ &  (3) \\
MACS0744              & 07:44:47.831 & +39:27:25.50 &  2.209  &     (4)         & 3.01$\pm$0.18           &  (6)\\
Sextet                        &13:11:26.466 & $-$01:19:56.28 &    3.042  &   (5)        & 4.43$\pm$0.33           & (13) \\
RXJ1347-11            & 13:47:29.271 & $-$11:45:39.47& 1.773  &     (6)         & 3.61$\pm$0.15           &  (6)\\
A1689-Blob              & 13:11:28.686 & $-$01:19:42.54 &  2.595  &     (6)        & 4.70$\pm$1.05      &  (13)\\
Cl0024                      & 00:26:34.407 & +17:09:54.97 & 1.679  &      (7)     & 1.38$\pm$0.15            &   (20)\\
MACS0025              & 00:25:27.686 & $-$12:22:11.23 &  2.378  &    (8)       & 1.75$\pm$0.25         &  (21) \\
MACS0451              & 04:51:57.186 & +00:06:14.87 &  2.013  &    (4)       & 4.22$\pm$0.27             &  (6) \\
MACS1423              & 14:23:50.775 & +24:04:57.45  & 2.530  &   (9)       &  1.10$\pm$0.12                 &  (9)\\
RXJ1053                  & 10:53:47.707 & +57:35:10.75 & 2.576  &   (9b)      & 4.03$\pm$0.12             &  Appendix A. \\
A1689-Highz           & 13:11:25.445 & $-$01:20:51.54 & 4.860  &  (10)  & 1.99$\pm$0.10             &  (13) \\
A2218-Ebbels         &  16:35:49.179 & +66:13:06.51 & 2.518  &  (11)  & 3.81$\pm$0.30               &  (12) \\
A2218-Flanking      & 16:35:50.475 & +66:13:06.38 & 2.518  &  (12)      & 2.76$\pm$0.21              &  (12) \\
MACS0712              &  07:12:17.534 & +59:32:14.96 & 2.646  &  (4)    & 3.60$\pm$0.32               &   (6) \\
Cl0949                      & 09:52:49.716 & +51:52:43.45 & 2.394  &  (13)       & 2.16$\pm$0.24              & (13)\\
A1835                       & 14:01:00.951 & +02:52:23.40& 2.071  &   (13)     & 4.50$\pm$0.32              &  (13) \\
A773                         & 09:17:57.403 & +51:43:46.57 & 2.300  &   (13)     & 2.69$\pm$0.35              & (13)\\
A2218-Mult              & 16:35:48.952 & +66:12:13.76 & 3.104  &    (12)    & 3.39$\pm$0.18              & (12)\\
A2218-Smm             &16:35:55.033 & +66:12:37.01  & 2.517  &   (14)  & 3.01$\pm$0.16               &  (12) \\
A68-C4                      & 00:37:07.716 & +09:09:06.44 &  2.622  &   (1)   & 4.15$\pm$0.16               & (13) \\
CL2244                     & 22:47:11.728 & $-$02:05:40.29 & 2.240 & (15)  & 4.08$\pm$0.30              & (15)\\
\hline
AC114-S2                 & 22:58:48.826 & $-$34:47:53.33 & 1.867  &   (16)    & $2.01\pm0.17$             & (22)\\
AC114-A2                 & 22:58:47.787 & $-$34:48:04.33 & 1.869  &    (16)    & $1.70\pm0.15$           & (22) \\
HORSESHOE           & 11:48:33.140 & +19:30:03.20 &   2.379  &   (17)  &   3.70$\pm$0.18             & (17) \\
CLONE                       & 12:06:02.090 & +51:42:29.52 &  2.001  &   (18)    & 3.62$\pm$0.12           &(24)\\
J0900+2234              &  09:00:02.790 & +22:34:03.60 & 2.032 & (23)   &   1.70$\pm$0.08             & (23) \\
\end{tabular}
\medskip\par

(1) \citet{Richard07} 
(2) \citet{Smail07} 
(3) \citet{Allam07} 
(4) \citet{Jones} 
(5) \citet{Frye07}  
(6) Richard et al. 2010c in preparation 
(7) \citet{Broadhurst00} 
(8) \citet{macs0025} 
(9) \citet{Limousin10}  
(9b) \citet{Hasinger}
(10) \citet{Frye02}  
(11) \citet{Ebbels} 
(12) \citet{Ardis} 
(13) \citet{Richard10a} 
(14) \citet{Kneib04b} 
(15) \citet{Mellier}  
(16) \citet{Lemoine03}  
(17) \citet{Belokurov} 
(18) \citet{Lin09} 
(19) \citet{Dye07} 
(20) Jauzac et al. in preparation 
(21) Smith et al. (2010) in preparation 
(22) \citet{Campusano}
(23) \citet{Bian}
(24) \cite{Jones10b}

\caption{\label{targets}The current sample of lensed galaxies. From left to right: astrometry, redshift and 
reference, magnification and reference. The sources below the line represent those drawn from the literature.}
\end{table*}
\clearpage

\begin{table}
\begin{tabular}{llll}
Run & Date & Seeing (\arcsec) & Photometric?\\
\hline
A   & 2005 October 13 & 0.5 & Photometric\\
B   & 2006 July 24 & 0.8 & Clear \\
C   & 2007 January 12 & 1.0 & Clear\\
D   & 2007 May 3 & 0.5-0.6 & Clear\\
E   & 2007 September 1 & 0.5 & Photometric\\
F   & 2008 March 23 & 0.4-0.5 & Photometric\\
G   & 2008 August 24 & 0.5-0.9 & Clear\\
\end{tabular}
\caption{\label{runs} NIRSPEC observing runs and conditions}
\end{table}

\begin{table*}

\begin{tabular}{lcccllllll}
ID             & $z$   & Runs & Filters   & [O{\sc ii}]& H$\beta$      & [O{\sc iii}]$\lambda5007$ & H$\alpha$ & [N{\sc ii}] & [S{\sc ii}]$^{(a)}$\\
\hline
A68-C1         & 1.583 & A    & N3        &            & $83.6\pm11.4$ & $273\pm17$    &            &             &\\
CEYE           & 3.074 & B    & N6        & $983\pm92$ & $490\pm22$    &$1030\pm32$    &            &             &\\
8OCLOCK        & 2.736 & C    & N4,N7     & $2690\pm66$&               &                 &$7450\pm90$ &$1140\pm37$&\\
MACS0744       & 2.209 & C    & N7        &            &               &                 &$1540\pm40$ &$461\pm23$   & $557\pm30$ \\
Sextet         & 3.042 & D    & N5,N7     & $<32$      & $200\pm20$    & $850\pm29.6$    &            &             &\\
RXJ1347-11     & 1.773 & D    & N1        &$1130\pm32$ &               &                 &            &             &\\
A1689-Blob     & 2.595 & D    & N7        &            &               &                 & $426\pm123$& $<39$       &\\
Cl0024         & 1.679 & E    & N5        &            &               &                 &$1140\pm34$ & $323\pm19$  &\\
MACS0025       & 2.378 & E    & N7        &            &               &                 &$87\pm10$   & $<66.7$     &\\
MACS0451       & 2.013 & E    & N6        &            & $1174\pm61$   & $7899\pm118.$   &$3170\pm59$ & $729\pm30$  & $999\pm45$ \\
MACS1423       & 2.530 & F    & N6,N7     &            & $123\pm40$    & $120\pm22$      & $313\pm39$ & $<45.8$     &\\
RXJ1053        & 2.576 & F    & N3,N6,N7  & $510\pm26$ & $624\pm32$    & $1980\pm44.5$   & $1640\pm42$& $213\pm18$  &\\
A1689-Highz    & 4.860 & F    & N7        & $43\pm17$  &               &                 &            &             &\\
A2218-Ebbels   & 2.518 & F    & N3,N6,N7  & $164\pm13$ & $108\pm11$    & $170\pm13$      & $273\pm18$ & $<21$       &\\
A2218-Flanking & 2.518 & F    & N6,N7     &            & $180\pm14$    & $402\pm20$      & $444\pm22$ & $<14$       &\\
MACS0712       & 2.646 & F    & N3,N6,N7  & $<32$      & $352\pm20$    & $719\pm40$      &$1070\pm34$ & $225\pm19$  &\\
Cl0949         & 2.394 & F    & N3,N5,N7  & $321\pm20$ & $<42.3$       & $819\pm30$      & $403\pm20$ & $<12$       &\\
A1835          & 2.071 & F    & N5        &            & $400\pm20$     & $879\pm46$     & $1700\pm150$ & $111\pm20$  &\\
A773           & 2.300 & F    & N6,N7     &            & $268\pm21$    & $108\pm11$      & $78\pm9$   & $126\pm12$  &\\
A2218-Mult     & 3.104 & G    & N4,N6     & $1310\pm40$& $871\pm31$    & $1920\pm70$     &            &             &\\
A2218-Smm      & 2.517 & G    & N6        &            & $688\pm29$    & $1030\pm34$     &            &             &\\
A68-C4         & 2.622 & G    & N6        &            & $97\pm15$     & $<22$           &            &             &\\
CL2244         & 2.239 & ISAAC    & J,H,K & $370\pm80$ & $<42$         & $67\pm10$       & $463\pm70$ & $<20$       &\\
\multicolumn{7}{l}{$^{(a)}$ Sum of the [S{\sc ii}] $\lambda$6717  and $\lambda$6731 line fluxes.}\\
\hline
\multicolumn{3}{l}{S/N Composite} & \multicolumn{4}{l}{[N{\sc ii}]$\lambda6584$/H$\alpha$=0.16$\pm$0.03, [S{\sc ii}]$\lambda\lambda6717,6731$/H$\alpha$=0.11$\pm$0.04 } \\
\multicolumn{3}{l}{Low M$_*$ Composite} & \multicolumn{4}{l}{[N{\sc ii}]$\lambda6584$/H$\alpha$=0.13$\pm$0.10}
\end{tabular}
\caption{\label{fluxes} Emission line measurements. Fluxes are given in units of 10$^{-19}$ ergs s$^{-1}$ cm$^{-2}$.}
\end{table*}

\begin{table*}
\begin{tabular}{llllllrrcl}
ID                &    $z$  & $r_{1/2}$        & $\sigma$         & M$_{dyn}$             & M$_{*}$               & SFR                & SFR$_{\rm corr}$&log(Z)                     & E(B-V)\\
                  &         & (kpc)        & (km/s)           & ($10^{10}$ M$_\odot$) & ($10^{10}$ M$_\odot$) &(M$_\odot\ yr^{-1})$&(M$_\odot\ yr^{-1})$&                           & \\
\hline
A68-C1            &  1.583  & 1.09$\pm$0.18& $72^{+16}_{-18}$ & $0.66^{+0.40}_{-0.27}$& $0.24_{-0.09}^{+0.03}$& 0.8$\pm$0.1        & 0.9$\pm$0.2     &                           & 0.02\\
CEYE              &  3.074  & 1.75$\pm$0.21& 54$\pm$4$^{(a)}$ & $0.59^{+0.16}_{-0.11}$& $5.74_{-0.90}^{+1.01}$& 37.6$\pm$4.3       & 77.3$\pm$8.8    &$8.64^{+0.12}_{-0.16}$     & 0.17\\
8OCLOCK           &  2.736  & 1.47$\pm$0.38& 45$\pm$5         & $0.35^{+0.16}_{-0.13}$& $1.78_{-0.86}^{+2.11}$& 98.0$\pm$28.       & 232.$\pm$48.    &$8.66^{+0.11}_{-0.12}$     & 0.22\\ 
MACS0744          &  2.209  & 1.00$\pm$0.22& $81^{+9}_{-9}$   & $0.76^{+0.33}_{-0.25}$& $0.99_{-0.19}^{+0.18}$&  6.6$\pm$1.0       & 11.9$\pm$1.8    &$8.91^{+0.13}_{-0.13}$     & 0.19\\
Sextet            &  3.042  & 0.30$\pm$0.10& $101^{+8}_{-8}$  & $0.35^{+0.17}_{-0.15}$& $0.02_{-0.01}^{+0.01}$&  1.1$\pm$0.3       & 2.4$\pm$0.7     &$8.00^{+0.44}_{-0.50}$     & 0.18\\
RXJ1347-11        &  1.773  & 2.48$\pm$0.27& $<46$            & $<0.61$               & $0.12_{-0.07}^{+0.02}$&  2.2$\pm$0.3       & 5.3$\pm$0.7     &                           & 0.16\\
A1689-Blob        &  2.595  &              &                  &                       & $0.01_{-0.01}^{+0.01}$&  0.6$\pm$0.4       &                 &                           & 0.0\\
Cl0024            &  1.679  & 10.0$\pm$1.2 & 69$\pm$5$^{(a)}$ & $5.52^{+1.45}_{-1.06}$& $2.32_{-0.35}^{+0.34}$& 14.6$\pm$1.9       & 34.5$\pm$4.6    &$8.89^{+0.13}_{-0.14}$     & 0.28\\
MACS0025          &  2.378  &              &                  &                       & $0.16_{-0.09}^{+0.05}$&  1.7$\pm$0.4       & 2.1$\pm$0.5     &                           & 0.08\\    
MACS0451          &  2.013  & 2.50$\pm$0.33& 80$\pm$5$^{(a)}$ & $1.86^{+0.47}_{-0.36}$& $1.36_{-0.58}^{+0.83}$&  3.4$\pm$0.8       & 5.9$\pm$1.3     &$8.80^{+0.13}_{-0.12}$     & 0.18\\
MACS1423          &  2.530  &              &                  &                       &                       &  8.5$\pm$1.4       &                 &                           & \\
RXJ1053           &  2.576  & 3.62$\pm$0.45& 68$^{+6}_{-6}$   & $1.94^{+0.58}_{-0.41}$& $0.42_{-0.08}^{+2.34}$&  3.8$\pm$0.4       & 90.5$\pm$2.3    &$8.68^{+0.11}_{-0.12}$     & 0.37\\
A1689-Highz       &  4.860  &              &                  &                       & $0.92_{-0.20}^{+0.24}$&  4.2$\pm$1.7       &                 &                           & 0.0\\
A2218-Ebbels      &  2.518  &              &                  &                       & $0.79_{-0.24}^{+0.29}$&  0.7$\pm$0.2       & 1.1$\pm$0.3     &$8.37^{+0.20}_{-0.20}$     & 0.16\\
A2218-Flanking    &  2.518  & 2.36$\pm$0.55& $50^{+17}_{-24}$ & $0.68^{+0.62}_{-0.48}$& $0.12_{-0.02}^{+0.03}$&  2.3$\pm$0.4       & 5.5$\pm$1.0     &                           & 0.28\\
MACS0712          &  2.646  & 0.75$\pm$0.23& $82^{+15}_{-17}$ & $0.58^{+0.39}_{-0.30}$& $2.89_{-2.88}^{+3.87}$&  3.0$\pm$0.7       & 5.6$\pm$1.4     &$8.77^{+0.14}_{-0.14}$     & 0.20\\
Cl0949            &  2.394  & 3.50$\pm$0.88& 66$\pm3^{(a)}$   & $1.77^{+0.60}_{-0.52}$& $1.54_{-1.02}^{+2.55}$&  3.6$\pm$0.7       & 7.5$\pm$1.5     &$8.10^{+0.06}_{-0.05}$     & 0.24\\
A1835             &  2.071  & 1.52$\pm$0.07& $124^{+39}_{-43}$& $2.71^{+1.82}_{-1.06}$& $0.06_{-0.02}^{+0.03}$&  1.1$\pm$0.3       & 2.5$\pm$0.6     &$8.40^{+0.40}_{-0.40}$     & 0.26\\
A773              &  2.300  & 0.38$\pm$0.05& $66^{+21}_{-29}$ & $0.19^{+0.15}_{-0.11}$& $0.15_{-0.05}^{+0.48}$&  0.4$\pm$0.1       & 0.9$\pm$0.3     &                           & 0.27\\
A2218-Mult        &  3.104  & 3.75$\pm$0.44& $57^{+21}_{-30}$ & $1.41^{+1.19}_{-0.90}$& $2.06_{-0.35}^{+0.38}$& 18.6$\pm$2.9       & 216$\pm$34      &$8.50^{+0.34}_{-0.23}$     & 0.58\\
A2218-Smm         &  2.517  & 1.86$\pm$0.40& $82^{+7}_{-7}$   & $1.45^{+0.56}_{-0.43}$& $1.84_{-0.26}^{+0.33}$& 10.1$\pm$1.4       & 21.7$\pm$3.1    &                           & 0.18\\
A68-C4            &  2.622  &              &                  &                       & $0.01_{-0.01}^{+0.01}$&  0.5$\pm$0.1       & 0.7$\pm$0.2     &                           & 0.05\\
CL2244            &  2.2399 &              &                  &                       & $0.13_{-0.04}^{+0.05}$&  1.4$\pm$0.4       & 2.1$\pm$0.6     &$8.42^{+0.48}_{-0.13}$     & 0.13\\
%
%
Composite 1       & $2.37\pm0.34$ & & &                                               & $0.97_{+3.37}^{-0.75}$ &                   &                 &$8.67^{+0.12}_{-0.12}$     &\\
Composite 2       & $2.52\pm0.10$ & & &                                               & $0.18_{+0.10}^{-0.10}$ &                   &                 &$8.64^{+0.19}_{-0.38}$     &\\
\hline\\
AC114-S2$^{(b)}$  &  1.867  &              &                  & 0.53$\pm$0.12         & $0.32_{-0.12}^{+0.10}$& 30.                &                 &8.78$\pm$0.20$^{(e)}$      & 0.30 \\
AC114-A2$^{(b)}$  &  1.869  &              &                  & 2.36$\pm$0.67         & $0.52_{-0.09}^{+0.31}$& 15.                &                 &8.94$\pm$0.20              & 0.40 \\
HORSESHOE$^{(c)}$ & 2.38    & 2.5          &                  & $1.0$                 &                       & $73.$              &                 &8.49$\pm$0.16              & 0.15 \\
CLONE$^{(c)}$     & 2.00    & 2.9          &                  & $2.2$                 &                       & $32.$              &                 &8.51$\pm$0.20              & 0.24 \\
J0900+2234$^{(d)}$& 2.032   &              &                  & 7.2                   & 0.6                   & 116$\pm$16         &                 &8.12$\pm$0.19              & 0.25 \\
\end{tabular}

$^{(a)}$\citet{Jones} $^{(b)}$\citet{Lemoine03} $^{(c)}$\citet{Hainline} $^{(d)}$\citet{Bian}

$^{(e)}$log(Z) value derived using the upper branch of the R23 diagram (see text for details).

\caption{\label{masses} Physical properties of the sample. From left to right: ID, redshift, magnification, half-light radius, measured velocity dispersion (corrected from instrumental resolution, see text for details), dynamical mass, stellar mass,  SFR, metallicity, stellar extinction from the SED fitting. The SFR and stellar masses are corrected for the lensing magnification factor, and include the aperture corrections. Average redshift and stellar mass are given for the composite spectra described in the text.}

\end{table*}

\label{lastpage}

\appendix

\section{RXJ1053 mass model}

\begin{figure}
\includegraphics[width=8.5cm,angle=0]{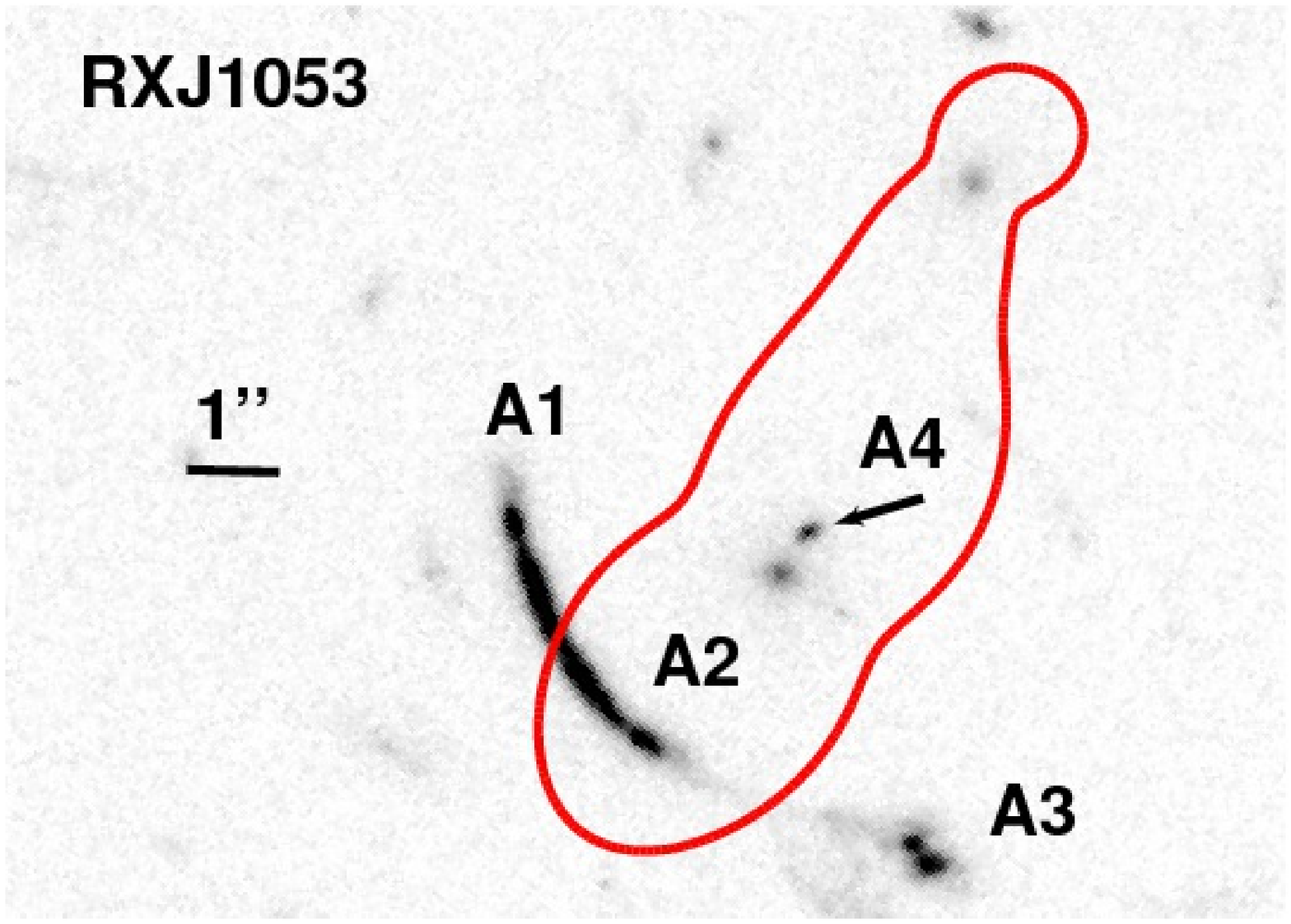}
\caption{\label{rxj1053} Central region around the BCG of the lensing cluster RXJ1053, showing the 
giant arc A visible as 4 images A1 to A4. The red line outlines the critical curve at the redshift $z=2.576$ of the 
source A.}
\end{figure}

In order to derive the magnification factor for the source at $z=2.576$ in the lensing cluster RXJ1053, 
we constructed a parametric mass model of the central region of the cluster using the Lenstool 
software \citep{Jullo}. Following similar strong-lensing works (e.g. \citealt{Richard09}), we assume 
the cluster mass distribution to follow a double Pseudo-Isothermal Elliptical (dPIE, \citealt{Ardis}) profile, and we add 
two central cluster members as lower-scale perturbations in the mass distribution. This model is constrained 
by  the location of 3 images of the giant arc, clearly detected thanks to their morphology and 
symmetry on the V-band HST image (Fig. \ref{rxj1053}). A fourth central image (A4) is predicted by the best 
fit model and identified on the same image. The best fit model has an rms $\sigma=0.15$'' between the predicted and 
observed positions of the 4 images. We report the best-fit parameters of the mass distribution in Table \ref{rxj1053table}.

\begin{table*}
\begin{tabular}{llllllll}
Component & $x$ & $y$ & $e$ & $\theta$ & $\sigma$ & r$_{\rm core}$ & r$_{\rm cut}$ \\
                      & ('') & ('') & & (deg.) & (km\ s$^{-1}$) & (kpc) & (kpc) \\                      
\hline
Cluster & $[0]$ & $[0]$ & 0.55$\pm$0.11 & 52.6$\pm$4.5 & 705$^{+169}_{-149}$ & 60$\pm$23 & 1000.0\\
BCG & $[0]$ & $[0]$  & [0.146] & [41.4] & 705$\pm$150 & [0.] & 39$^{+40}_{-5}$\\
Gal1 & [2.1] & [4.2] & [0.16] & [42.4] & 202$\pm$87 & [0.] & 20$^{+30}_{-3}$ \\
\end{tabular}
\caption{\label{rxj1053table} Best fit parameters for the mass distribution reproducing the multiple components of the 
giant arc in RXJ1053. The dPIE parameters are given for each component: centre, ellipticity and orientation, velocity dispersion, core and cut radii.}
\end{table*}

\newpage
\begin{landscape}
\begin{table}
\caption{\label{SEDtable}Multi-walength photometry (AB magnitudes) of the lensed sources in our sample. First column gives the UV $\beta$ slope (see text for 
details). We use the notation BB'GVRII'JH as a shortcut 
for the various HST broad-band filters. Filter names in italics refer to ground-based images.}
\begin{tabular}{llllllllllllll}
ID & $\beta$ & F450W(B)  & F555W(G) & F775W(I')& F850LP & F110W/$J$ & F160W/$H$ & $K$/{\it Ks} & IRAC 3.6$\mu$m & IRAC 4.5$\mu$m\\
   &         & F475W(B') & F606W(V) & F814W(I) &        &           &           &              &                &               \\
   &         &        & F702W(R) &          &        &           &           &              &                &               \\
\hline
8OCLOCK & 1.60$\pm$0.15 &
B 21.94$\pm$0.10 &
V 21.36$\pm$0.10  &
I 21.10$\pm$0.10  & &
J 21.12$\pm$0.10  &
H 20.77$\pm$0.10  & &
   20.16$\pm$0.12  &
   19.84$\pm$0.12 \\
A68-C1 & [2.0] & &
R 24.02$\pm$0.04 & & &
$J$ 22.41$\pm$0.08 &
$H$ 22.30$\pm$0.08 &
$K$ 21.77$\pm$0.22 &
   21.98$\pm$0.15 &
   22.18$\pm$0.15 \\
COSMICEYE  & 0.38$\pm$0.11 & &
V 20.54$\pm$0.02 & 
I 20.01$\pm$0.05 & &
J 20.07$\pm$0.05 & 
H 19.32$\pm$0.05 & 
$K$       18.82$\pm$0.10 & 
          18.26$\pm$0.15 & 
          18.31$\pm$0.15 \\
MACS0744    & 1.30$\pm$0.07& &
G 23.04$\pm$0.03 &
I 22.73$\pm$0.03 & & & &
{\it Ks} 21.22$\pm$0.24 &
    20.45$\pm$0.08 &
    20.42$\pm$0.16 \\
A1689-Sextet  & 1.77$\pm$0.19&
B' 23.25$\pm$0.05 &
V 22.36$\pm$0.05 &
I' 22.29$\pm$0.05 &
  22.34$\pm$0.05 &
J 22.36$\pm$0.05 &
$H$      21.94$\pm$0.05 &
$K$      21.89$\pm$0.05 &
   21.87$\pm$0.05 &
   22.02$\pm$0.05
\\
RXJ1347       & 1.65$\pm$0.10 &
B' 21.61$\pm$0.06 & &
I 21.41$\pm$0.06 & &
$J$ 21.16$\pm$0.18  &
$H$ 20.52$\pm$0.29 &
$K$ 20.56$\pm$0.25 &
          20.30$\pm$0.17 &
          20.40$\pm$0.11 \\
A1689-Blob    & 2.83$\pm$0.15 &
B' 23.13$\pm$0.05 &
V 23.14$\pm$0.05 &
I' 23.25$\pm$0.05 &
  23.45$\pm$0.05 &
J 23.65$\pm$0.05 &
$H$ 23.28$\pm$0.05 &
$K$ 23.44$\pm$0.05&
    22.50$\pm$0.05&
    22.67$\pm$0.05
\\
Cl0024        & 1.04$\pm$0.21 & 
B' 21.78$\pm$0.06 &
V 21.53$\pm$0.06&
I' 21.32$\pm$0.06&
  21.10$\pm$0.07&
J 20.59$\pm$0.08&
H 20.27$\pm$0.08&
$K$     20.12$\pm$0.12&
          19.56$\pm$0.10&
          19.58$\pm$0.10 \\
MACS0025     & 1.77$\pm$0.15 & &
G 23.70$\pm$0.05 &
I 23.60$\pm$0.08 & & & & &
          22.94$\pm$0.17 &
          22.69$\pm$0.13 \\
MACS0451     & 1.61$\pm$0.21 & &
V 19.60$\pm$0.07 &
I 19.47$\pm$0.07 & & & & &
          17.77$\pm$0.16 &
          17.80$\pm$0.16  \\
MACS1423        & 2.43$\pm$0.05& &
V 23.78$\pm$0.06 &
I 23.70$\pm$0.06 &  & & &
{\it Ks}      $>23.08$ \\
RXJ1053         & [2.0] & &
V 21.25$\pm$0.05 & & & &
H 20.01$\pm$0.10 \\
A1689-Highz  & 3.32$\pm$0.23 & &
V 24.93$\pm$0.05 &
I' 23.23$\pm$0.05 &
  23.20$\pm$0.05 &
J 23.31$\pm$0.05 & 
$H$ 23.66$\pm$0.12& 
$K$ 22.99$\pm$0.08&
   22.81$\pm$0.12&
\\
A2218-Ebbels    & 1.89$\pm$0.15 & 
B' 21.24$\pm$0.11 &
V 20.75$\pm$0.06  &
I' 20.66$\pm$0.06 &
  20.71$\pm$0.06 &
J 20.44$\pm$0.08&
H 19.78 $\pm$0.08 &
$K$ 19.79$\pm$0.12&
          19.41$\pm$0.16&
          19.39$\pm$0.16  \\
A2218-Flanking  & 1.86$\pm$0.13 & 
B' 23.32$\pm$0.05 &
V  22.70$\pm$0.05 &
I' 22.61$\pm$0.05 &
   22.65$\pm$0.05 &
J  22.31$\pm$0.08 &
H  22.08$\pm$0.09 &
$K$  22.17$\pm$0.14 & &\\ 
MACS0712      & $1.17\pm0.15$ & &
V 21.93$\pm$0.05 &
I  21.65$\pm$0.05 \\
Cl0949        & [2.0] & &
V  22.09$\pm$0.05 & & & & & & 
          20.13$\pm$0.20  \\
A1835         & $1.76\pm0.15$ & &
R 21.35$\pm$0.07 & &
   21.28$\pm$0.05 &
$J$      20.84$\pm$0.07 &
H 20.87$\pm$0.04 &
$K$      20.51$\pm$0.08 \\
A773          & [2.0] & & 
R 22.67$\pm$0.05 & & & & & & 
          21.26$\pm$0.12   &
          21.10$\pm$0.14 \\
A2218-Mult    & 1.22$\pm$0.14 &
B' 24.95$\pm$0.16 &
V  22.23$\pm$0.06&
I' 21.95$\pm$0.05&
   21.93$\pm$0.05& & &
$K$     19.91$\pm$0.13&
          18.93$\pm$0.14&
          18.74$\pm$0.13 \\
A2218-Smm      & 1.52$\pm$0.11 & 
B' 23.52$\pm$0.05 &
V 23.12$\pm$0.05&
I' 23.13$\pm$0.05&
  22.97$\pm$0.04&
J 22.65 $\pm$0.07&
H 21.88$\pm$0.11&
$K$ 21.29$\pm$0.14&
          20.45$\pm$0.12 &
          20.09$\pm$0.13 \\
A68-C4        & [2.0] & & 
R  23.31$\pm$0.04 & & & 
$J$      22.91$\pm$0.29&
$H$      22.98$\pm$0.35& &
          23.02$\pm$0.12&
          23.21$\pm$0.15 \\
CL2244        & 1.76$\pm$0.15 & &
G 21.35$\pm$0.06 &
I 21.25$\pm$0.06  & & 
$J$      20.74$\pm$0.18 &
$H$      20.49$\pm$0.18 & & 
          20.53$\pm$0.12 & 
          20.34$\pm$0.12 \\
AC114-A2       & 0.98$\pm$0.17 & &
R  22.16$\pm$0.06 & &
   21.86$\pm$0.05&
$J$      21.19$\pm$0.18&
$H$      21.30$\pm$0.11&
$K$      20.92$\pm$0.13 \\
AC114-S2       & 1.25$\pm$0.17 & &
R 22.96$\pm$0.06 & &
  22.74$\pm$0.04 &
$J$      22.33$\pm$0.07 &
$H$      22.21$\pm$0.06 &
$K$      22.04$\pm$0.07 &
          21.63$\pm$0.12 &
          21.51$\pm$0.08 \\
\end{tabular}
\end{table}
\end{landscape}

\end{document}